\documentclass[12pt]{iopart}
\usepackage[dvips]{graphicx}%
\usepackage{bm}%
\usepackage{cite}
\usepackage{multirow}

\usepackage{iopams}
\expandafter\let\csname equation*\endcsname\relax
\expandafter\let\csname endequation*\endcsname\relax
\usepackage[fleqn]{amsmath}
\usepackage{etoolbox}
\usepackage[normalem]{ulem} 
\usepackage[colorlinks=true,linkcolor=blue,citecolor=blue ,urlcolor=blue]{hyperref}

\begin{document}

\title[Geometry, quantum correlations, and phase transitions]{Geometry, quantum correlations, and phase transitions in the $\Lambda$-atomic configuration}
\author{O. Casta\~nos, S. Cordero, R. L\'opez-Pe\~na, and E. Nahmad-Achar}
\eads{\mailto{ocasta@nucleares.unam.mx}}

\address{Instituto de Ciencias Nucleares, Universidad Nacional
Aut\'onoma de M\'exico, Apdo. Postal 70-543 M\'exico 04510 CDMX.}

\begin{abstract}
The quantum phase diagram for a finite $3$-level system in the $\Lambda$ configuration, interacting with a two-mode electromagnetic field in a cavity, is determined by means of information measures such as fidelity, fidelity susceptibility and entanglement, applied to the reduced density matrix of the matter sector of the system. The quantum phases are explained by emphasizing the spontaneous symmetry breaking along the separatrix. Additionally, a description of the reduced density matrix of one atom in terms of a simplex allows a geometric representation of the entanglement and purity properties of the system. These concepts are calculated for both, the symmetry-adapted variational coherent states and the numerical diagonalisation of the Hamiltonian, and compared. The differences in purity and entanglement obtained in both calculations can be explained and visualised by means of this simplex representation. 
\end{abstract}

\maketitle

\section{Introduction}

Particularly interesting are the phase transitions at zero temperature, which are due to competing ground state phases.  These transitions have relevance in condensed matter because they help to understand many systems, such as two-dimensional electron gases, high-temperature superconductors, and magnetic insulators, together with new states of matter such as topological insulators and superconductors~\cite{bitko96, QI2011}. In quantum optics they play a fundamental role in explaining the transition from radiant to super-radiant regimes in models of matter interacting with electromagnetic modes in a cavity~\cite{hepp73, greentree06}. In molecular physics, there are signatures of quantum phase transitions in the bending excitations of triatomic molecules in algebraic model Hamiltonians~\cite{iachello11,bernal13}.  In many of these examples there are potential applications to the design of quantum technologies and quantum critical metrology~\cite{sachdev11, zanardi08, garbe20}. 
Information measures such as fidelity, fidelity susceptibility, Bures distance and entanglement, have been useful tools in establishing the phase diagram of a physical system, not only to determine the localization of quantum phase transitions for a finite number of particles, but to obtain information on their scaling behavior~\cite{you07, castanos12, nahmad-achar13}.

In this work, the measures of fidelity and fidelity susceptibility are used on the reduced density matrix of the matter sector to determine the quantum phase diagram, as well as the variational phase diagram associated to {\it symmetry-adapted states} (SAS), in a {\it finite} system. An important contribution is the use of a simplex representation to understand the differences between these SAS states and the numerical calculation of the direct diagonalisation of the Hamiltonian. We have previously demonstrated that the points in the vecinity of the separatrix (the loci of minimal critical points, i.e., the points where the ground state of the system suffers significant changes) have maximum entanglement properties between the matter and field Hilbert spaces; here we show that the same result is obtained by studying the single-atom correlation properties. At the separatrix one can observe a spontaneous symmetry breaking; for example, the change in the ground state at the transition from the normal to a collective regime, is seen as a dominance of only a few basis components in the state to that of many basis components, while the first-order transition between collective sectors involves a change of symmetry from the algebra of one subspace to the algebra of another subspace. On this separatrix the occupation probabilities of the second and third energy levels take the same value, viz. $1/2$.

This paper is organised as follows: In section~\ref{qpd} a review of the Hamiltonian model of the three different atomic configurations of $3$-level systems is presented together with their corresponding variational phase diagrams in the limit $N_a \to \infty$. Section~\ref{s.lambda} discusses the symmetries of the Hamiltonian to be restored in the variational test SAS states for the $\Lambda$ configuration. In section~\ref{s.fidelity} the measures of fidelity and fidelity susceptibility are used on the reduced density matrix of the matter to determine the quantum phase diagram associated to the SAS states. The single-atom correlations and entanglement properties, together with the comparison between the SAS states and the exact solution from the diagonalisation of the Hamiltonian, are discussed in section~\ref{s.oneatom}. It is shown that in the vicinity of the separatrix one finds states with maximum entanglement properties. Finally, in Sec.~\ref{s.conclusions}  concluding remarks and a summary of the results are given.

\section{Review of quantum phase diagrams}
\label{qpd}

The Hamiltonian describing a system of $N_a$ atoms of $3$ levels interacting dipolarly with $2$ modes of an electromagnetic field in a resonator is given by~\cite{cordero15}
\begin{eqnarray}
\label{eq.H}
\bm{H} = \sum^2_{s=1}\Omega_{s}\, \bm{a}_{s}^\dag\, \bm{a}_{s}  + \sum^{3}_{k=1} \,\omega_k \, \bm{A}_{kk}
 + \bm{H}_{int}\,;
 \end{eqnarray}
the first two terms are the diagonal contributions of field and matter subsystems, respectively, with $\bm{a}_{s}^\dag$ and $\bm{a}_{s}$ the creation and annihilation operators of the mode with frequency $\Omega_s$ ($s=1,2$), the atomic energy levels are $\hbar\, \omega_k$ ($k=1,2,3$) where we have taken $\hbar=1$, and $\bm{A}_{kk}$ are the collective matter weight operators.

\begin{figure}
\begin{center}
\includegraphics[width=0.85\linewidth]{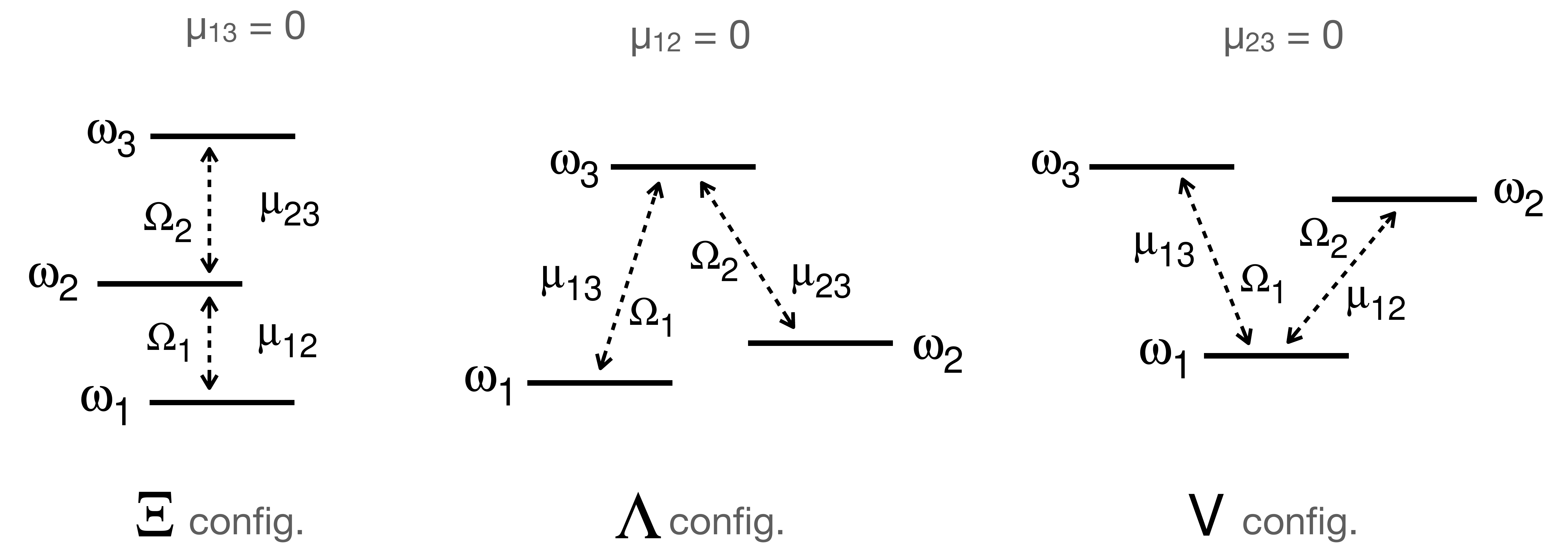}
\caption{Schematic diagrams for the distinct $3$-level atomic configurations, showing the dipolar couplings $\mu_{ij}$ between the atomic levels and the field modes.} 
\label{AtomicConfig}
\end{center}
\end{figure}

The interaction Hamiltonian between the field and matter sectors is given by
\begin{equation}
\bm{H}_{int} = -\frac{1}{\sqrt{N_a}} \sum_{j<k}^3 \sum_{s=1}^2 \mu_{jk}^{(s)} \left(\bm{A}_{jk} + \bm{A}_{kj}\right) \left(\bm{a}_s^\dagger + \bm{a}_s \right)
\label{hint}
\end{equation}
where $\mu_{jk}^{(s)}$ is the field-matter dipole coupling constant of the atomic levels $j$, $k$ with the field mode $s$, and the collective atomic transition operators $\bm{A}_{kj}$ obey a unitary algebra in three dimensions u$(3)$ satisfying the commutation relations
\begin{equation}
\left[\bm{A}_{lm},\bm{A}_{kj}\right] = \delta_{mk}\,\bm{A}_{lj}-\delta_{jl}\,\bm{A}_{km}\, .
\label{eq.commAij}
\end{equation}
For $N_a$ identical bosonic particles, the first $\bm{C}_1$ and second $\bm{C}_2$ order Casimir operators  of u$(3)$ are given by
\begin{equation}
\bm{C}_1 = \sum_{k=1}^3 \bm{A}_{kk} = \bm{N}_a \, , \qquad \bm{C}_2 = \sum_{k=1}^3 \sum_{j=1}^3 \bm{A}_{jk} \bm{A}_{kj} =  \bm{N}_a (\bm{N}_a+2) \, ,
\end{equation} 
and are constants of motion. In our model, each electromagnetic field mode only promotes  transitions between one pair of atomic levels, and the atomic levels are labelled with frequencies satisfying the inequalities $\omega_1<\omega_2<\omega_{3}$. Consequently, each atomic configuration, $\Xi$, $\Lambda$ and $V$, is determined by the two coupling constants $\mu_{jk}^{(s)}$ which do not vanish. Hereafter we will take  $(\mu^{(1)}_{12},\,\mu^{(2)}_{23}) = (\mu_{12},\,\mu_{23})$ for  $\Xi$; $(\mu^{(1)}_{13},\,\mu^{(2)}_{23}) = (\mu_{13},\,\mu_{23})$ for $\Lambda$; and $(\mu^{(1)}_{12},\,\mu^{(2)}_{13}) = (\mu_{12},\,\mu_{13})$ for $V$ (cf. Fig.~\ref{AtomicConfig}).
%

\subsection*{Case $N_a \to\infty$}

For the model Hamiltonian in (\ref{eq.H}) the quantum phase diagram in the case $N_a \to \infty$ may be determined using  the following procedure~\cite{gilmore93, cordero15}: 
\begin{enumerate}
	
\item 
A tensorial product of coherent states of u$(3)$ for the matter and Heisenberg-Weyl coherent states for the field is selected as a test function, 
\begin{equation}
\vert N_a;\vec{\alpha};\, \vec{\gamma}\rangle := |\vec{\alpha}\rangle \otimes |N_a; \,\vec{\gamma}\rangle\, ,
\end{equation}
with arbitrary complex vectors \[\vec{\gamma} = (\gamma_1,\,\gamma_2,\,\gamma_3) = \left(\rho_1 e^{i\phi_1},\,\rho_2 e^{i\phi_2},\,\rho_3 e^{i\phi_3}\right)\,,\quad \vec{\alpha} =(\alpha_1,\alpha_2)=\sqrt{N_a}\left(r_1 e^{i\theta_1},\,r_2 e^{i\theta_2}\right)\,,\] for the u(3) and field sectors, respectively.

The coherent state of the matter sector is (for $N_a$ atoms)~\cite{iachello80, ginocchio80, iachello87}, 
\begin{equation}\label{eq.coh.matter}
|N_a;\, \vec{\gamma}\rangle =\frac{1}{\sqrt{N_a!}}\,  \frac{(\gamma_1\,\bm{b}_1^\dag + \gamma_2\,\bm{b}_2^\dag + \gamma_3\,\bm{b}_3^\dag)^{N_a}}{\left(|\gamma_1|^2 + |\gamma_2|^2 + |\gamma_3|^2\right)^{N_a/2}} \, |0\rangle_{M},
\end{equation} 
where the vacuum state in the Fock basis is defined by $\bm{b}_k \, |0\rangle_{M}=0$, ($k=1,2,3$) while, for the electromagnetic field, the Glauber-Sudarshan-Klauder coherent states are used~\cite{klauder85}: 
\begin{equation}
 |\vec{\alpha}\rangle = e^{-\left(\frac{\vert \alpha_1\vert^2}{2}+\frac{\vert \alpha_2\vert^2}{2}\right)} \sum^\infty_{k_1,k_2=0} \frac{\alpha_1^{k_1}}{\sqrt{k_1!}}\,  \frac{\alpha_2^{k_2}}{\sqrt{k_2!}} \, \vert k_1\, k_2\, \rangle \, ,
\end{equation}
with $\vert k_1\, k_2\rangle$ denoting Fock states with $k_1$ and $k_2$ photons of the first and second modes, respectively.

\item
The expectation value of the Hamiltonian in these test states yields the so-called variational energy surface as a function of the state variables which, normalised per particle,  takes the form
\begin{equation}
{\cal E}(\vec{\alpha}, \vec{\alpha}^{\, *} \,; \vec{\gamma}, \vec{\gamma}^{\,*}) = \frac{1}{N_a} \, \langle N_a;\vec{\alpha};\, \vec{\gamma} \, \vert \bm{H}\vert N_a;\vec{\alpha};\, \vec{\gamma}\rangle \, ,
\label{supE}
\end{equation}
and is a real function of $\vec{\gamma}, \vec{\gamma}^{\,*}, \vec{\alpha}, \vec{\alpha}^{\, *}$ and the parameters of $\bm{H}$ (the dipolar coupling  strengths, the frequencies of the atomic levels, and those of the field modes); without loss of generality one may take $\gamma_1=1$ (cf. Eq.~(\ref{eq.coh.matter})).

\item
The critical points of~(\ref{supE}) are calculated, and the {\it minimum} energy surface is obtained from them. The advantage of using these test functions is that one can get analytical solutions, and we find that all the complex variables appearing in the variational test wave function are real at the critical points, i.e., the critical phases are $\phi_k^{(c)} = 0,\,\pi; \ \theta_s^{(c)} = 0,\,\pi$ and hence  $\gamma^{(c)}_k=\pm \rho^{(c)}_k$ ($k=1,2,3$), and $\alpha^{(c)}_s=\pm \sqrt{N_a} \, r^{(c)}_s$ ($s=1,2$), the correct set of critical phases must yield an attractive matter-field interaction in the minimum energy surface. One finds~\cite{cordero15} (in order to simplify the notation we have written $\rho^{(c)}_k \equiv \rho_k,\ (k=2,3)$)
\begin{eqnarray}
(r_1,r_2)^{(c)}_V&=& \left\{ (0,\, 0)_N, \, \left(\frac{2\, \mu_{12} \, \rho_2}{\Big(1+ \rho^{2}_2\Big) \,\Omega_1}, \, 0\right)_{S_{12}}, \, \left(0, \frac{2\, \mu_{13} \, \rho_3}{\Big(1+\rho^{2}_3\Big) \,\Omega_2}\right)_{S_{13}} \right\} \, , \nonumber \\
(r_1,r_2)^{(c)}_\Lambda&=& \left\{ (0,\, 0)_N,\, \left(\frac{2\, \mu_{13} \, \rho_3}{\Big(1+\rho^{2}_3\Big) \,\Omega_1},\, 0\right)_{S_{13}}, \,\left(0,\, \frac{2\, \mu_{23} \,  \rho_3}{\Big(1+  \rho^{2}_3\Big)\, \Omega_2} \right)_{S_{23}}\right\} \, ,\nonumber \\
(r_1,r_2)^{(c)}_\Xi&=& \left\{ (0,\, 0)_N, \, \left(\frac{2\, \mu_{12} \, \rho_2}{\Big(1+ \rho^{2}_2\Big) \,\Omega_1} ,\, 0 \right)_{S_{12}},\, \left(0,\, \frac{2\, \mu_{23} \,   \rho_3}{\Big(1+  \rho^{2}_3 \Big)\,\Omega_2}\right)_{S_{23}} \right\}\, .
\label{campo}
\end{eqnarray}
The expressions above show that the polychromatic phase diagram divides itself into monochromatic regions: $N$ for the {\it normal} region of independent decay, and $S_{jk}$ for the {\it collective} or {\it superradiant} region of coherent decay corresponding to the atomic levels $(j,k)$. The critical points $\rho_k$ are given in Table~\ref{tpc} for each region in the $\Xi$-, $\Lambda$- and $V$-atomic configurations where $\mu^{(s)}_{jk} \neq 0$. 

Notice that in Table~\ref{tpc} the region $S_{13}$ appears twice; this is because both configurations $\Lambda$ and $V$ have regions with the same label, $S_{13}$, but associated to different field modes: 
$\vert \alpha_{1},\, 0\rangle$ in the $\Lambda$ configuration, and $\vert 0\, , \alpha_{2} \rangle$ in the $V$ configuration.

\end{enumerate}

\begin{table}
\caption{Critical points of the minimum energy surface in each region, for the different atomic configurations. We have defined the Bohr frequencies $\omega_{jk}=\omega_j-\omega_k$.}
\begin{center}
\begin{tabular}{c||c|c|c|c|c|}
 Region &  \hspace{1mm} $\rho^{(c)}_1$  \hspace{1mm} &$\rho^{(c)}_2$ & $\rho^{(c)}_3$ & Conditions  \\[2mm]
\hline
\hline
&&& &\\[-3mm]
$N$ & \, 1 \,& 0 &  0 &  \\[3mm]
$S_{12}$ & 1 &  \hspace{1mm} $\displaystyle\sqrt{\frac{4\, \mu^2_{12} - \Omega_1\, \omega_{21}}{4\, \mu^2_{12} + \Omega_1\, \omega_{21}}}$  \hspace{1mm}& 0 &  \hspace{1mm} $4\, \mu^2_{12} \geq \Omega_1 \, \omega_{21} $  \hspace{1mm} \\[7mm] 
$S_{23}$ & \, 0 \, & 1 
 &  \hspace{1mm} $\displaystyle \sqrt{\frac{4\, \mu^2_{23} - \Omega_2\, \omega_{32}}{4\, \mu^2_{23} + \Omega_2\, \omega_{32}}}$  \hspace{1mm} & $4\, \mu^2_{23} \geq  \Omega_2\, \omega_{32} $\\[7mm] 
$S_{13}(\Lambda)$ &\, 1\, & 0&  $\displaystyle\sqrt{\frac{4\, \mu^2_{13} - \Omega_1\, \omega_{31}}{4\, \mu^2_{13} + \Omega_1\, \omega_{31}}}$ & $4\, \mu^2_{13} \geq \Omega_1\, \omega_{31}$ \\[7mm] 
$S_{13}(V)$ &\, 1\, & 0&  $\displaystyle\sqrt{\frac{4\, \mu^2_{13} - \Omega_2\, \omega_{31}}{4\, \mu^2_{13} + \Omega_2\, \omega_{31}}}$ & $4\, \mu^2_{13} \geq \Omega_2\, \omega_{31}$ \\[5mm] 
\hline
\end{tabular}
\end{center}
\label{tpc}
\end{table}

The quantum phase diagrams of the $3$-level system interacting with two modes of electromagnetic radiation present the following regions (see Table~\ref{twfe}): 
\begin{itemize}
\item[i)]
The normal region, denoted by $N$, where there are no photons present and all the atoms occupy their ground state level. Thus the energy takes the constant value $\hbar\, \omega_1$.
\item[ii)]
The collective region $S_{jk}$, where there are photons of that mode of the electromagnetic field which promotes transitions between the atomic levels $\omega_j \rightleftharpoons \omega_k$.
\end{itemize}

The boundaries between regions (i.e., the geometric loci of the separatrix) are obtained by taking the set of algebraic equations associated to their energies,
\begin{eqnarray}
{\cal E}_N &=& {\cal E}_{S_{12}} ,  \, \, \quad {\cal E}_N = {\cal E}_{S_{13}}, \, \, \quad {\cal E}_N = {\cal E}_{S_{23}} , \nonumber 
\\
 {\cal E}_{S_{12}} &=& {\cal E}_{S_{13}},\quad  {\cal E}_{S_{12}} = {\cal E}_{S_{23}}, \quad  {\cal E}_{S_{13}} = {\cal E}_{S_{23}} \, ,
\label{sepeq}
\end{eqnarray}
whose explicit expressions are given in the last column of Table~\ref{twfe}.
\begin{table}[htp]
\caption{Variational ground state and corresponding energy in each region of the quantum phase diagram; $\alpha^{(c)}_s = \sqrt{N_a}\,r^{(c)}_s,\ (s=1,2)$, with $r^{(c)}_s$ given by eq.~(\ref{campo}), and in order to simplify the notation we have written $\rho^{(c)}_k \equiv \rho_k,\ (k=2,3)$, which are given in Table~\ref{tpc} for each of the appropriate region.}
\begin{center}
\begin{tabular}{c||c|c|c|c|c|}
 Region & Ground State & ${\cal E}$ \\[2mm]
\hline
\hline
&& \\[-1mm]
$N$ & $\displaystyle\vert 0, \,0\rangle\otimes \frac{(\bm{b}_1^\dag)^{N_a}}{\sqrt{N_a!}} \, |0\rangle_{M}$ & $\omega_1$\\[3mm]
$S_{12}$&  \hspace{1mm} $\displaystyle\vert \alpha^{(c)}_{1}\, , 0\rangle \otimes  \frac{(\bm{b}_1^\dag + \rho_2\,\bm{b}_2^\dag )^{N_a}}{\sqrt{N_a!\, \left(1 + \rho^2_2 \right)^{N_a}}} \, |0\rangle_{M}$  \hspace{1mm} &  \hspace{1mm}$\displaystyle-\frac{\mu^2_{12}}{\Omega_1} -\frac{\Omega_1(\omega_2-\omega_1)^2}{16\, \mu^2_{12}}+\frac{\omega_1+\omega_2}{2}$  \hspace{1mm}\\[7mm] 
$S_{23}$ & $\displaystyle\vert 0\, ,\alpha^{(c)}_{2} \rangle \otimes  \frac{(\bm{b}_2^\dag + \rho_3\,\bm{b}_3^\dag )^{N_a}}{\sqrt{N_a!\, \left(1 + \rho_3^2 \right)^{N_a}}} \, |0 \, \rangle_{M}$ & $\displaystyle-\frac{\mu^2_{23}}{\Omega_2} -\frac{\Omega_2(\omega_3-\omega_2)^2}{16\, \mu^2_{23}}+\frac{\omega_2+\omega_3}{2}$\\[7mm] 
$S_{13}(\Lambda)$ &   $\displaystyle\vert \alpha^{(c)}_{1} \,, 0 \rangle \otimes  \frac{(\bm{b}_1^\dag + \rho_3\,\bm{b}_3^\dag )^{N_a}}{\sqrt{N_a!\, \left(1 + \rho^2_3 \right)^{N_a}}} \, |0\rangle_{M}$ & $\displaystyle-\frac{\mu^2_{13}}{\Omega_1} -\frac{\Omega_1(\omega_3-\omega_1)^2}{16\, \mu^2_{13}}+\frac{\omega_1+\omega_3}{2}$\\[7mm]
$S_{13}(V)$ &   $\displaystyle\vert 0,\,\alpha^{(c)}_{2} \rangle \otimes  \frac{(\bm{b}_1^\dag + \rho_3\,\bm{b}_3^\dag )^{N_a}}{\sqrt{N_a!\, \left(1 + \rho^2_3 \right)^{N_a}}} \, |0\rangle_{M}$ & $\displaystyle-\frac{\mu^2_{13}}{\Omega_2} -\frac{\Omega_2(\omega_3-\omega_1)^2}{16\, \mu^2_{13}}+\frac{\omega_1+\omega_3}{2}$\\[7mm]
\hline
\end{tabular}
\end{center}
\label{twfe}
\end{table}

Very significant changes in the composition of the ground state take place when one crosses the regions given in Table~\ref{tpc} of the energy surface. The phase diagram in the finite case is much richer than that obtained in the case $N_a\to\infty$,  because in general one finds more transitions~\cite{lopez-pena21}. As the number of particles $N_a$ increases, these new transitions deform into those that remain in the limit $N_a\to\infty$. We have called these ``phase transitions'', in contrast  to crossovers, precursors, or other names sometimes given in the literature~\cite{Brandes2003}, since, physically, the ground state suffers considerable changes in its structure, as seen by the mix of basis states that enter into play when minute fluctuations in any parameter of the system are had near of separatrix in the exact quantum solution. Thus physically, the ground state acquires a different composition and properties~\cite{cordero21}.

\subsubsection*{Atomic $V$-configuration\\[3mm]}

\noindent
Using equations~(\ref{sepeq}) for the atomic $V$-configuration, we find the following boundaries:
%
\numparts

	\begin{eqnarray}
0 \leq \mu_{12} \leq \frac{1}{2} \sqrt{\Omega_1 \, \omega_{21}}\quad  &\land& \quad  \mu_{13} = \frac{1}{2} \sqrt{\Omega_2 \, \omega_{31}}\,, \label{sepVa}\\
0 \leq \mu_{13} \leq \frac{1}{2} \sqrt{\Omega_2 \, \omega_{31}}\quad  &\land& \quad \mu_{12} = \frac{1}{2} \sqrt{\Omega_1 \, \omega_{21}}\,,
\label{sepVb}
	\end{eqnarray}
for the transition between regions $N$ and $S_{12}$, and $N$ and $S_{13}$, respectively.  According to the Ehrenfest classification, the order of the phase transition is given by the lowest derivative of the energy surface with respect to the variables $\rho_k$ for which at the critical points a discontinuity is had. In our case,
one finds that both transtions are of second order.

Note that the equalities in Eqs.~(\ref{sepVa}) and (\ref{sepVb}) determine the critical coupling strenghts for the corresponding two-level system $\omega_j,\ \omega_k$, i.e., the dipolar transitions $\omega_1 \rightleftharpoons \omega_2$ for the mode $\Omega_1$ and $\omega_1 \rightleftharpoons \omega_3$ for the mode $\Omega_2$; denoting these points by $\bar{\mu}_{jk}^{(s)}= \sqrt{\Omega_s \, \omega_{kj}}/2$, we define dimensionless coupling strenghts as 
\begin{equation*}
x_{12}=\frac{\mu_{12}}{\bar{\mu}_{12}^{(1)}} \,, \quad x_{13}=\frac{\mu_{13}}{\bar{\mu}_{13}^{(2)}} \,.\end{equation*}
The separatrix for the transition between collective regions $S_{12}$ and $S_{13}$ may then be written as
\begin{eqnarray}
 x_{13}^2 &=& 1+\frac{\omega_{21}}{\omega_{31}}\frac{x_{12}^2-1}{2x_{12}^2}\left[ x_{12}^2-1+f_V(x_{12};\omega_{21},\omega_{31}) \right]  \quad \land \quad x_{12}\geq1\,,
\end{eqnarray}
\endnumparts
%
with
\begin{equation*}
f_V(x_{12};\omega_{21}, \omega_{31})=\sqrt{ (x_{12}^2 - 1)^2 + 4x_{12}^2\frac{\omega_{31}}{\omega_{21}} }\,.
\end{equation*}
Between collective regions a first order transition is obtained, due to the fact that the basis of the variational ground state changes suddenly from the subspace ${\cal H}_{12}$ to the subspace ${\cal H}_{13}$, where the subsystems coexist.

\subsubsection*{Atomic $\Xi$-configuration\\[3mm]}

\noindent
In this case one finds the following boundaries:
%
\numparts
\label{sepX}
	\begin{eqnarray}
0 \leq \mu_{12} \leq \frac{1}{2} \sqrt{\Omega_1 \, \omega_{21}}\quad  &\land& \quad  \mu_{23} = \frac{1}{2} \sqrt{\Omega_2} \, \left( \sqrt{\omega_{21}}+\sqrt{\omega_{31}} \right)\,,  \\
0 \leq \mu_{23} \leq \frac{1}{2} \sqrt{\Omega_2} \, \left( \sqrt{\omega_{21}}+\sqrt{\omega_{31}} \right)\quad  &\land& \quad \mu_{12} = \frac{1}{2} \sqrt{\Omega_1 \, \omega_{21}}\,,
\label{sepXa}
	\end{eqnarray}
for the transition between regions $N$ and $S_{12}$ (second order), and $N$ and $S_{23}$ (first order), respectively. The dimensionless coupling strengths are
\begin{equation*}
x_{12}=\frac{\mu_{12}}{\bar{\mu}_{12}^{(1)}} \,, \quad x_{23}=\frac{\mu_{23}}{\bar{\mu}_{23}^{(2)}} \,, 
\end{equation*}
and the separatrix for the transition between collective regions $S_{12}$ and $S_{23}$ is given by
\begin{eqnarray}
 x_{23}^2 &=& 1+\frac{\omega_{21}}{\omega_{32}}\frac{x_{12}^2+1}{2x_{12}^2}\left[ x_{12}^2+1+f_\Xi(x_{12};\omega_{21},\omega_{32}) \right]  \quad \land \quad x_{12}\geq1\,,
\end{eqnarray}
\endnumparts
%
with
\begin{equation*}
f_\Xi(x_{12};\omega_{21}, \omega_{32})=\sqrt{ (x_{12}^2 +1)^2 + 4x_{12}^2\frac{\omega_{32}}{\omega_{21}} }\,,
\end{equation*}
which also corresponds to a first order transition, and denotes a phase coexistence locus.

\subsubsection*{$\Lambda$-atomic configuration\\[3mm]}

\noindent
Similarly for the atomic $\Lambda$ configuration the boundaries are:
%
\numparts
\label{sepL}
	\begin{eqnarray}
0 \leq \mu_{13} \leq \frac{1}{2} \sqrt{\Omega_1 \, \omega_{31}}\quad  &\land& \quad  \mu_{23} = \frac{1}{2} \sqrt{\Omega_2} \, \left( \sqrt{\omega_{31}}+\sqrt{\omega_{21}} \right)\,, \\
0 \leq \mu_{23} \leq \frac{1}{2} \sqrt{\Omega_2} \, \left( \sqrt{\omega_{31}}+\sqrt{\omega_{21}} \right)\quad  &\land& \quad \mu_{13} = \frac{1}{2} \sqrt{\Omega_1 \, \omega_{31}}\,,
\label{sepLa}
	\end{eqnarray}
for the transition between regions $N$ and $S_{13}$ (second order), and $N$ and $S_{23}$ (first order), respectively. The dimensionless coupling strengths take the form
\begin{equation*}
x_{13}=\frac{\mu_{13}}{\bar{\mu}_{13}^{(1)}} \,, \quad x_{23}=\frac{\mu_{23}}{\bar{\mu}_{23}^{(2)}} \,, 
\end{equation*}
and the separatrix for the transition between collective regions $S_{13}$ and $S_{23}$ (first order) is given by
\begin{eqnarray}
 x_{23}^2 &=& -1+\frac{\omega_{31}}{\omega_{32}}\frac{x_{13}^2+1}{2x_{13}^2}\left[ x_{13}^2+1+f_\Lambda(x_{13};\omega_{31},\omega_{32}) \right]  \quad \land \quad x_{13}\geq1\,,
\end{eqnarray}
\endnumparts
%
with
\begin{equation*}
f_\Lambda(x_{13};\omega_{31}, \omega_{32})=\sqrt{ (x_{13}^2 +1)^2 - 4x_{13}^2\frac{\omega_{32}}{\omega_{31}} }\,.
\end{equation*}

\begin{figure}
\begin{center}
\includegraphics[width=0.49\linewidth]{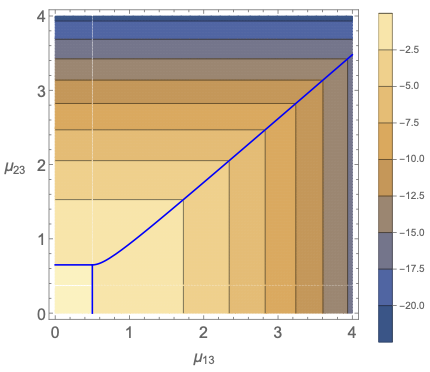}
\caption{Equipotentials of the energy surface and separatrix (continuous blue line) for the $\Lambda$-atomic configuration. The parameters are: $\omega_1=0$, $\omega_2=1/5$, $\omega_3=1$ for the atomic levels, and $\Omega_1=1$, $\Omega_2=4/5$ for the electromagnetic field. The coexistence  point of the phases $S_{N},\, S_{13}$ and $S_{23}$ is  given by $(\mu_{13},\mu_{23}) = (1/2,(5+\sqrt{5})/25)$. As the matter-field coupling increases, the energy of the ground state diminishes, as the dipolar interaction in the Hamiltonian is atractive.}
\label{qpdiagrams}
\end{center}
\end{figure}

As an example, in Fig.~\ref{qpdiagrams} we show equipotentials of the variational energy surface for the $\Lambda$-atomic configuration, together with the separatrix as a continuous (blue) line,  where we take (all along this work) the parameters: $\omega_1=0$, $\omega_2=1/10$, $\omega_3=1$ for the atomic levels, and $\Omega_1=1$, $\Omega_2=9/10$ for the electromagnetic field.

A similar behaviour of the quantum phase diagram is exhibited for the other atomic configurations, and from here on we focus on the $\Lambda$-atomic configuration because a similar procedure to construct the  SAS states can be  used for the other atomic configurations~\cite{cordero13b, cordero15,lopez-pena21}. 
%

\section{Symmetry-adapted coherent states for the $\Lambda$-atomic configuration}
\label{s.lambda}

The use of coherent states allows us to obtain analytic expressions for the difrerent observables of the system, which are helpful to study the whole of the Hilbert space including the asymptotic behaviours. Having said that, the coherent states do not respect the symmetries of the Hamiltonian; we may, however, project them using the parity transformations under which the Hamiltonian is invariant. The result of this are the {\it symmetry-adapted} states, which give a better description of the system and a very good prediction of the matter-field entanglement.

The interaction term of the Hamiltonian~(\ref{hint}) reads
\begin{eqnarray}
\label{eq.HL}
\bm{H}^{\Lambda}_{int}&=& - \frac{\mu_{13}}{\sqrt{N_a}} \left(\bm{A}_{13}+\bm{A}_{31}\right)\left(\bm{a}_{1} + \bm{a}_{1}^\dag\right) - \frac{\mu_{23}}{\sqrt{N_a}}\left(\bm{A}_{23}+\bm{A}_{32}\right)\left(\bm{a}_{2} + \bm{a}_{2}^\dag\right)\,. 
\end{eqnarray}
This Hamiltonian is invariant under the parity transformations~\cite{cordero15}
\begin{equation}
\bm{\Pi}_1 = e^{i \, \pi \bm{M}} \, , \quad \bm{\Pi}_2 = e^{i \, \pi \bm{K}}  \, ,
\end{equation}
with
\begin{equation}
\bm{M} = \bm{\nu}_1 +\bm{\nu}_2 + \bm{A}_{33}  \, ,  \quad \bm{K} = \bm{\nu}_2 + \bm{A}_{11} + \bm{A}_{33} \, ,
\end{equation}
where $\bm{\nu}_s$ is the photon number operator of mode $\Omega_s$ of the electromagnetic field.  
Both $\bm{M}$ and $\bm{K}$  are constants of motion when the rotating wave approximation (RWA) is considered.

Therefore the Hilbert space divides itself naturally into four subspaces, according to the even (${\rm e}$) or odd (${\rm o}$) parity of the eigenvalues of the operators $\bm{M}$ and $\bm{K}$: ${\cal H}:= {\cal H}_{\rm ee} \oplus {\cal H}_{\rm eo} \oplus {\cal H}_{\rm oe} \oplus {\cal H}_{\rm oo}$. 
In order to restore the symmetries of the Hamiltonian in the variational states, we project 
\begin{eqnarray}
&&\vert N_a; \alpha_1,\alpha_2;\gamma_1,\gamma_2,\gamma_3 \rangle_{\rm par(M)\,par(K)} = \nonumber \\
&& \phantom{xxxxxxxxxx} = {\cal N} \, \left( 1 + (-1)^{\rm par(M)}\bm{\Pi}_1\right) \left(1 + (-1)^{\rm par(K)}\bm{\Pi}_2 \right) \, | N_a; \alpha_1, \alpha_2; \gamma_1, \gamma_2, \gamma_3 \rangle \, ,
\end{eqnarray}
where ${\cal N}$ is the normalization constant; thus the following SAS states are obtained:
%
\numparts
\begin{eqnarray}
\fl
\vert N_a; \alpha_1,\alpha_2;\gamma_1,\gamma_2,\gamma_3 \rangle_{\rm ee} &=& 
{\cal N} \left( | N_a; \alpha_1, \alpha_2; \gamma_1,\gamma_2,\gamma_3 \rangle + 
| N_a; -\alpha_1, -\alpha_2; \gamma_1,  \gamma_2,  -\gamma_3 \rangle\right. \\
&&+ \left. | N_a; \alpha_1, -\alpha_2; -\gamma_1, \gamma_2 , -\gamma_3 \rangle + 
| N_a; -\alpha_1, \alpha_2; -\gamma_1, \gamma_2, \gamma_3 \rangle \right) \, ,\nonumber \\[3mm]
\fl
\vert N_a; \alpha_1,\alpha_2;\gamma_1,\gamma_2,\gamma_3 \rangle_{\rm eo} &=& {\cal N} \left( 
| N_a; \alpha_1, \alpha_2; \gamma_1,\gamma_2,\gamma_3 \rangle + 
| N_a; -\alpha_1, -\alpha_2; \gamma_1,  \gamma_2,  -\gamma_3 \rangle\right. \\
&&- \left. | N_a; \alpha_1, -\alpha_2; -\gamma_1, \gamma_2 , -\gamma_3 \rangle - 
| N_a; -\alpha_1, \alpha_2; -\gamma_1, \gamma_2, \gamma_3 \rangle \right) \, , \nonumber \\[3mm]
\fl
\vert N_a; \alpha_1,\alpha_2;\gamma_1,\gamma_2,\gamma_3 \rangle_{\rm oe} &=& {\cal N} \left( 
| N_a; \alpha_1, \alpha_2; \gamma_1,\gamma_2,\gamma_3 \rangle - 
| N_a; -\alpha_1, -\alpha_2; \gamma_1,  \gamma_2,  -\gamma_3 \rangle\right. \\
&&+ \left. | N_a; \alpha_1, -\alpha_2; -\gamma_1, \gamma_2 , -\gamma_3 \rangle - 
| N_a; -\alpha_1, \alpha_2; -\gamma_1, \gamma_2, \gamma_3 \rangle \right) \, ,\nonumber \\[3mm]
\fl
\vert N_a; \alpha_1,\alpha_2;\gamma_1,\gamma_2,\gamma_3 \rangle_{\rm oo} &=& {\cal N} \left( 
| N_a; \alpha_1, \alpha_2; \gamma_1,\gamma_2,\gamma_3 \rangle - 
| N_a; -\alpha_1, -\alpha_2; \gamma_1,  \gamma_2,  -\gamma_3 \rangle\right. \\
&&- \left. | N_a; \alpha_1, -\alpha_2; -\gamma_1, \gamma_2 , -\gamma_3 \rangle + 
| N_a; -\alpha_1, \alpha_2; -\gamma_1, \gamma_2, \gamma_3 \rangle \right) \, . \nonumber
\end{eqnarray}
\endnumparts
%
Notice that the SAS states are linear combinations of the mean field solutions and now the matter and field sectors are not separable. This is proved in the next sections.

The energy surfaces $\{ {\cal E}_{\rm ee}, {\cal E}_{\rm eo} , {\cal E}_{\rm oe}, {\cal E}_{\rm oo}\}$ associated to each one of the Hilbert subspaces is calculated,  by following the same procedure described above, and compared in order to find the ground state in each region of the control parameter space. These will be used to compute quantum information properties in the next section, and hence to learn about the characteristics of the quantum phase diagram of the system.

\section{SAS Quantum Phase Diagram}
\label{s.fidelity}

In order to find the loci of quantum phase transitions in parameter space we use concepts from quantum information, in particular, the fidelity and the fidelity susceptibility between neighbouring states.  A detailed description to acomplish this procedure is given in~\cite{cordero21,lopez-pena21}.

For pure states $\vert \Psi_1\rangle$ and $\vert \Psi_2\rangle$ the fidelity is defined by
\begin{equation}
F=|\langle \Psi_1|\Psi_2\rangle|^2 \, ,
\end{equation}
which measures the transition probability from one state to the other; it also denotes, in information theory, the similarity between two quantum states.  $F=1$ indicates that the two states are identical except for a global phase, whereas $F=0$ implies that the states are orthogonal.

For mixed states given by density operators $\bm{\rho}_1$ and $\bm{\rho}_2$, the fidelity measure is given by~\cite{uhlmann76,jozsa94,bengtsson17}
\begin{equation}
F(\bm{\rho}_1,\bm{\rho}_2) =\left\{ {\rm Tr}\left (\sqrt{ \sqrt{\bm{\rho}_1} \, \bm{\rho}_2\, \sqrt{\bm{\rho}_1}}\right) \right\}^2 \, ,
\label{fid-gen}
\end{equation}
which reduces to the previous expression for pure states, and which satisfies the relations
\begin{equation*}
0\leq F(\bm{\rho}_1,\bm{\rho}_2) \leq 1 , \quad F(\bm{\rho}_1,\bm{\rho}_2)=F(\bm{\rho}_2,\bm{\rho}_1),\quad 
 F(\bm{U} \bm{\rho}_1\bm{U}^\dag,\bm{U}\bm{\rho}_2\bm{U}^\dag) =F(\bm{\rho}_1,\bm{\rho}_2) \, ,
\end{equation*}
for any unitary operator $\bm{U}$ on the state space.

The fidelity susceptibility may be calculated through the expression
\begin{equation}
\chi(\bm{\rho}(x)) = 2 \, \frac{1-F(\bm{\rho}(x), \bm{\rho}(x+\delta x)) }{||\delta x||^2} \, ,\quad  \vert\vert\delta x\vert\vert \ll 1
\label{fid-sus}
\end{equation}
which is a more sensitive measure~\cite{gu10}, and where here $x$ and $x+\delta x$ denote the set of parameters characterizing neighbouring states.

Yet another way to determine the quantum phase diagram for a finite number of particles is by means of the Bures distance~\cite{bengtsson17}, which measures how far two probability densities are. It is closely related to the fidelity defined by Uhlmann and the R\'enyi relative entropy, and it is given by
\begin{equation}
d_B(\bm{\rho}_1,\bm{\rho}_2) = \sqrt{2- 2\sqrt{F(\bm{\rho}_1,\bm{\rho}_2)}} \, ,
\end{equation}
whose value lies in the interval $0\leq d_B \leq \sqrt{2}$. When $d_B=\sqrt{2}$ the quantum states are completely different, while for $d_B=0$ they are indistinguishable. Therefore, one can calculate the surface of maximum Bures distance in order to find the significant changes in the composition of the ground state.

Here we calculate the fidelity of the reduced density matrix of matter $\bm{\rho}_M$ between neighbouring states, as a function of the dimensionless dipolar coupling strengths using~(\ref{fid-gen}),
\begin{equation}
F(\bm{\rho}_M(x),\bm{\rho}_M(x+\delta x)) =\left\{ {\rm Tr}\left (\sqrt{ \sqrt{\bm{\rho}_M(x)} \, \bm{\rho}_M(x+\delta x)\, \sqrt{\bm{\rho}_M(x)}}\right) \right\}^2 \, ,
\end{equation}
with $x=(x_{13},x_{23})$, and we determine the surface of minimum fidelity by comparing the three following cases: 
\begin{eqnarray*}
&& F (\bm{\rho}_M(x_{13},x_{23}), \bm{\rho}_M(x_{13} +\delta x_{13},x_{23})) \, , \\
&& F (\bm{\rho}_M(x_{13},x_{23}), \bm{\rho}_M(x_{13} ,x_{23} +\delta x_{23}))\, , \\
&& F (\bm{\rho}_M(x_{13},x_{23}), \bm{\rho}_M(x_{13}+\delta x_{13} ,x_{23} +\delta x_{23})) \, .
\end{eqnarray*}
After the surface of minimum fidelity is determined, one evaluates the corresponding maximum surface of fidelity susceptibility defined by~(\ref{fid-sus}).  
\begin{figure}
\begin{center}
\includegraphics[width=0.39\linewidth]{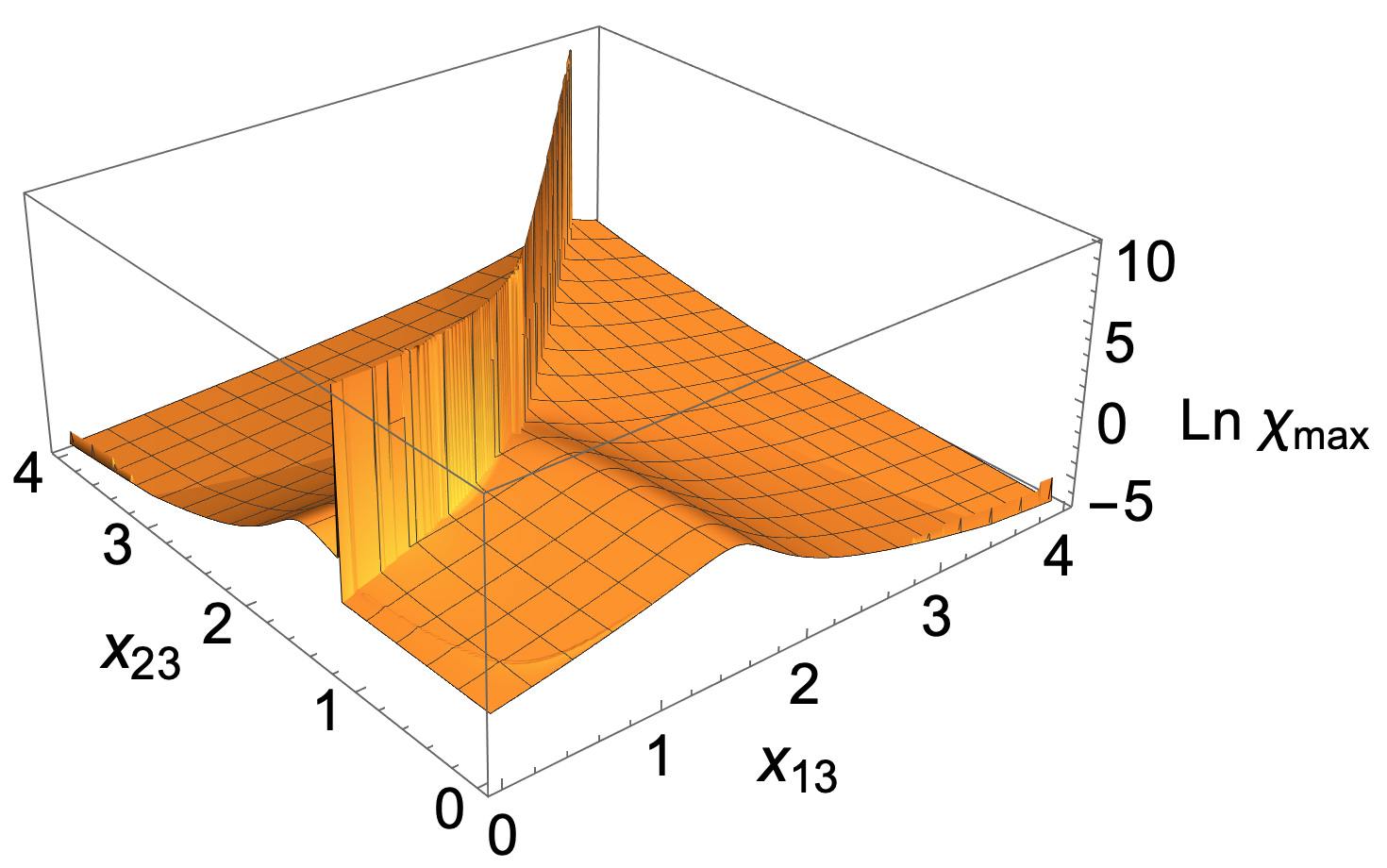}\hspace{2cm}
\includegraphics[width=0.29\linewidth]{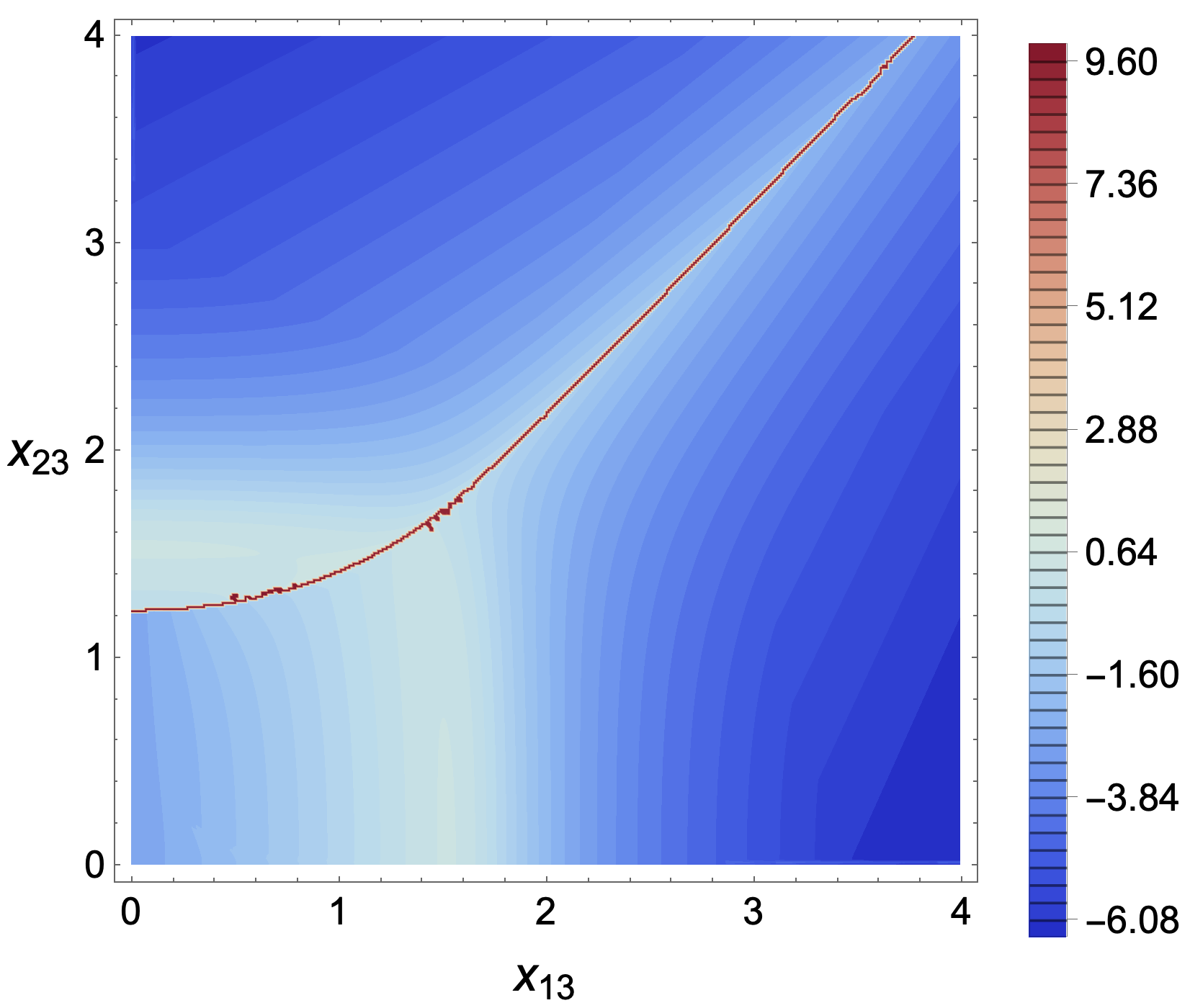}\\
\includegraphics[width=0.42\linewidth]{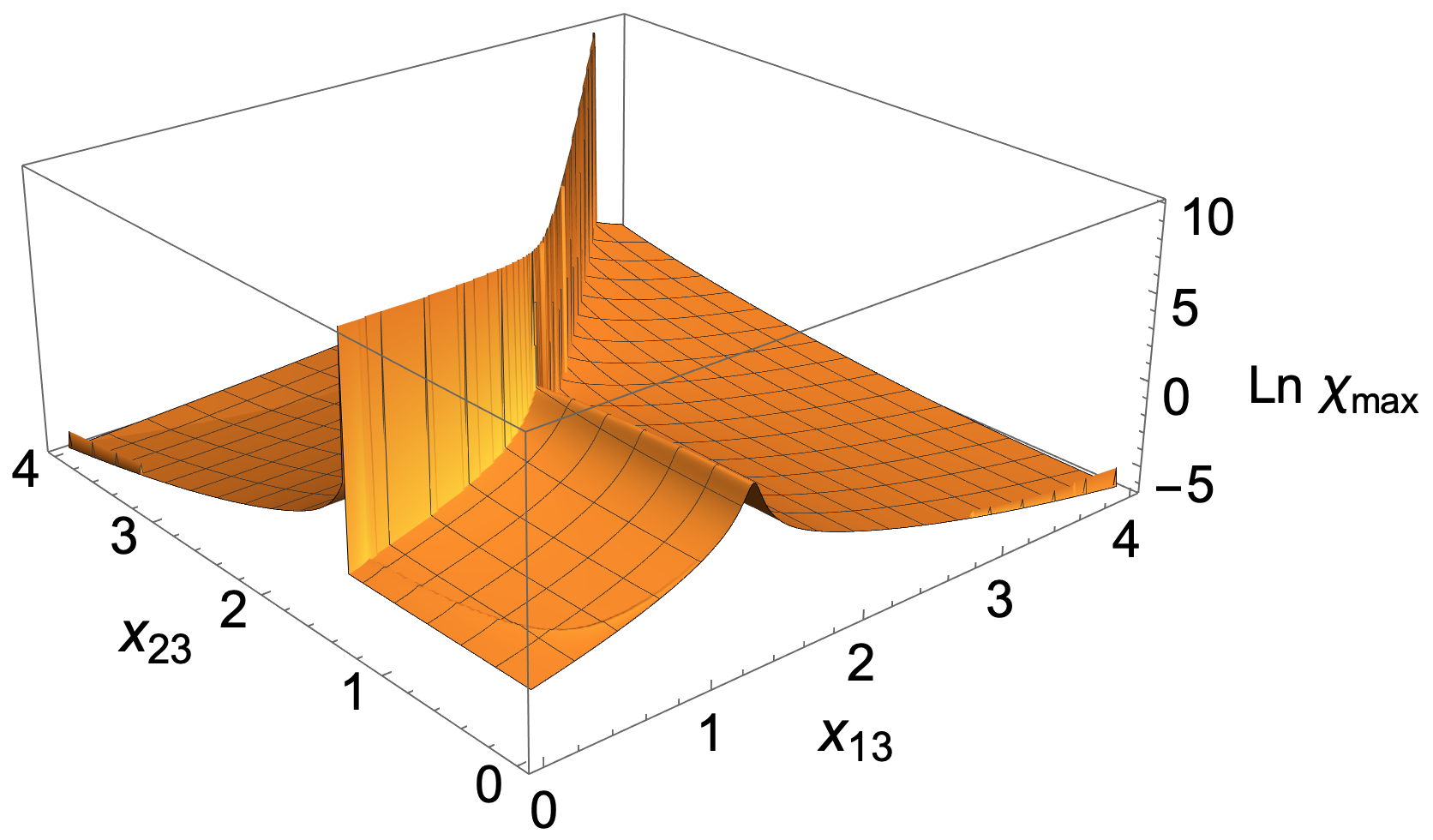}\hspace{1.6cm} 
\includegraphics[width=0.29\linewidth]{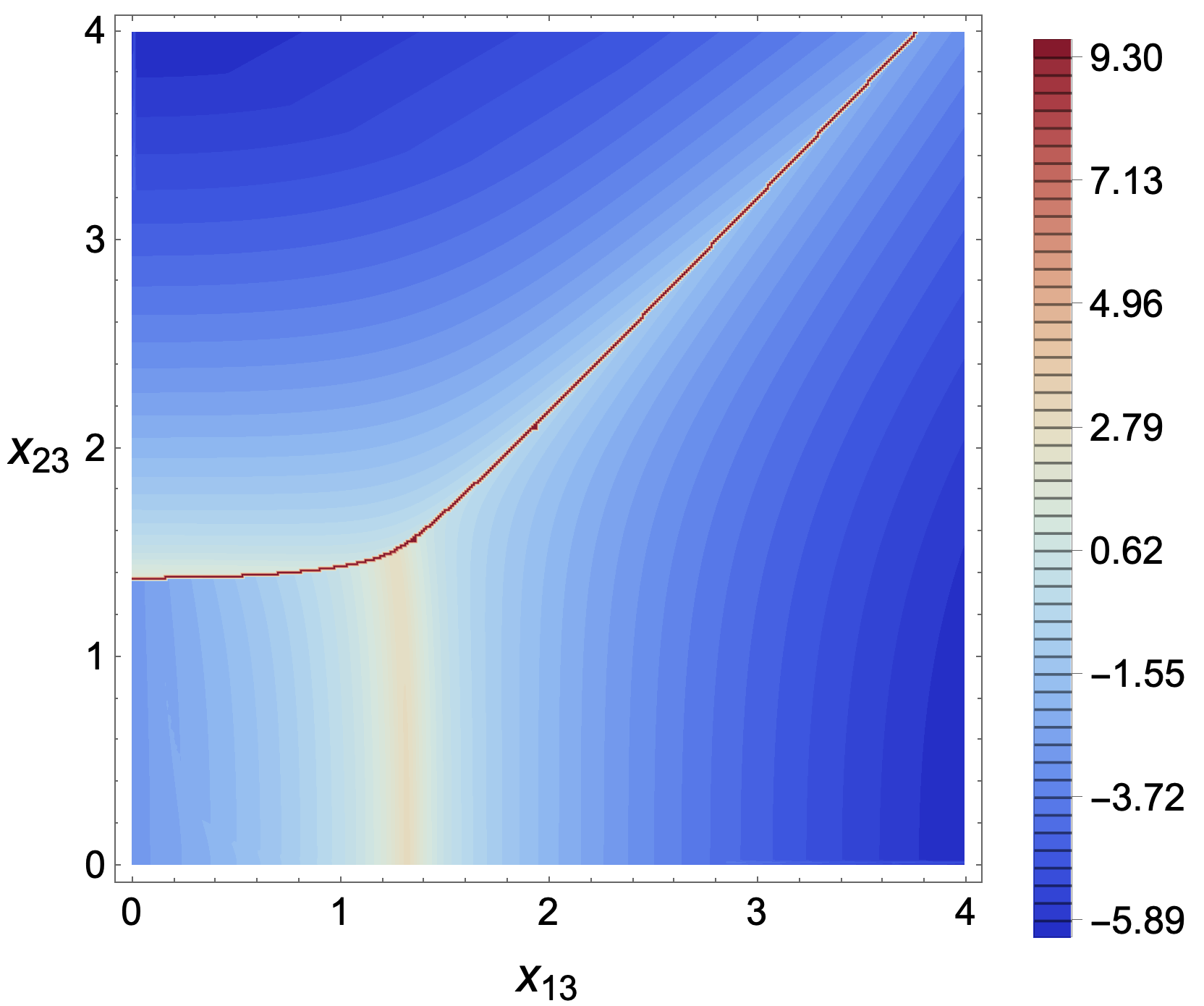}
\caption{Left: Quantum phase diagrams for $N_a=2$ (top) and $N_a=4$ (bottom) in the $\Lambda$-atomic configuration, obtained from the fidelity susceptibility by plotting the $\ln \chi$ as function of $x_{13}$ and $x_{23}$ for the even-even SAS variational function. Right: Corresponding contour plots. The parameters are the same as in Fig.~\ref{qpdiagrams}. Note that the local maxima of the surface of maximum susceptibility define the quantum phase diagram, that is, the separatrix of the Hamiltonian system.}
\label{qpBures}
\end{center}
\end{figure}

The reduced density matrix of the matter gives information about the quantum phase diagram of the complete system, as shown in Fig.~\ref{qpBures} for the $\Lambda$-atomic configuration, with $N_a=2$ and $N_a=4$ particles (cf. also Fig.~\ref{qpdiagrams}), where we plot the natural logarithm of the maximum fidelity susceptibility surface as a function of $x_{13}$ and $x_{23}$ for the SAS variational ground state. One can identify the normal (closest to the origin) and collective (far from the origin) regions.  The normal region is characterised by the atoms being mainly in their lowest energy state, that is, the occupation probability per particle of the lower atomic level is ${\cal P}_1 \approx 1$. The collective region is divided into two parts, the sector where there is a dominance of the subsystem $\mathcal{S}_{13} = \rm{span}\{\vert\nu_1,\,0;\, n_1,\,0,\,n_3\rangle\}$, and that with predominance of $\mathcal{S}_{23} = \rm{span}\{\vert 0,\,\nu_2;\, 0,\,n_2,\,n_3\rangle\}$. We then witness a spontaneous symmetry breaking from the ${\rm su}(2)_{13}$ algebra to the ${\rm su}(2)_{23}$ algebra.

\section{Quantum Correlations and Entanglement}
\label{s.oneatom}

For $m$ identical particles the expectation values of one- and two-body operators can be written in terms of the one- and two-particle mean values. In general, it is well known that the expectation value of a $k$-body operator in an $m$ particle space can be given by
\begin{equation}
\langle \hat{\cal O}(k) \rangle^{(m)} = \frac{m!}{(m-k)! \, k!} \langle   \hat{\cal O}(k) \rangle^{(k)} \, ,
\label{qce1}
\end{equation}
where $ \langle   \hat{\cal O} \rangle^{(k)}$  denotes the expectation value in the $k$-particle space~\cite{wong86}. Therefore a mixed particle rank operator $\hat{\cal O}$ of maximum rank 2 can be decomposed as
\begin{equation*}
\hat{\cal O} = \sum^2_{k=1} \hat{\cal O}(k) \, . 
\end{equation*}
Using~(\ref{qce1}) one has 
\begin{equation}
\langle \hat{\cal O}\rangle^{(m)} =m \, \langle   \hat{\cal O}(1) \rangle^{(1)}  + \frac{m(m-1)}{2}  \langle   \hat{\cal O}(2) \rangle^{(2)}\, .
\end{equation}

The generators $\bm{A}_{jk}$ of an algebra u$(3)$ are one body operators and, for a totally symmetric representation, take the form
\[
\bm{A}_{jk} = \bm{b}^\dagger_j \, \bm{b}_k   \quad  j,k = 1,2,3 \, ,
\]
where $\bm{b}_k$ and $\bm{b}^\dagger_j$ are annihilation and creation boson operators, respectively.  Thus their expectation values in an $N_a$-particle space satisfy the relation
\begin{equation}
\langle \bm{A}_{jk} \rangle^{(N_a)} = N_a \, \langle \bm{A}_{jk} \rangle^{(1)} . 
\label{qce2}
\end{equation}

Additionally, for the one-particle case the u$(3)$ generators can be realised by $3\times 3$ matrices,
\begin{equation}
(\bm{A}_{jk})_{\alpha\beta} = \Bigg\{
\begin{array}{cc}
 1 & \qquad \hbox{for} \, \alpha=j, \beta=k \\
 0 &\hbox{otherwise}
\end{array}
\end{equation}
Therefore, the expectation value of $\bm{A}_{jk}$ for the one-particle case determines the matrix element of the corresponding density matrix,
\begin{equation}
\langle \bm{A}_{jk} \rangle^{(1)} = \rho_{kj} \, . 
\end{equation}
In conclusion one can write the one-particle reduced density matrix $\bm{\rho}^{(1)}$ in terms of the expectation values in the Hilbert space for  $N_a$ particles, 
\begin{equation}
\bm{\rho}^{(1)}=\frac{1}{N_a} \, \left( 
\begin{array}{ccc}
\langle \bm{A}_{11}\rangle  &  \langle \bm{A}_{21} \rangle &  \langle \bm{A}_{31} \rangle \\
\langle \bm{A}_{12} \rangle  & \langle \bm{A}_{22} \rangle  &  \langle \bm{A}_{32} \rangle \\
\langle \bm{A}_{13} \rangle  &  \langle \bm{A}_{23} \rangle &   \langle \bm{A}_{33} \rangle
\end{array}
\right) \, ,
\label{qce3}
\end{equation}
where we simplify the notation $\langle \bm{A}_{jk} \rangle^{(N_a)} \to \langle \bm{A}_{jk} \rangle$, that is, the matrix elements of the one-particle reduced density matrix are determined in terms of the expectation values of the u$(3)$ generators in the Hilbert space for the $N_a$ particle system.  

For the considered Hamiltonian system the one-particle reduced density matrix is diagonal,  $\langle \bm{A}_{jk} \rangle =0$ for $j\neq k$, and these diagonal elements determine the occupation probabilities ${\cal P}_k$ for each atomic level, 
\begin{equation}
\bm{\rho}^{(1)} = {\rm diag}\,\left({\cal P}_1,{\cal P}_2,{\cal P}_3\right) =\frac{1}{N_a} \, {\rm diag}\, \left( \langle \bm{A}_{11}\rangle\, ,\, \langle \bm{A}_{22}\rangle\, ,  \, \langle \bm{A}_{33}\rangle \right) \, .
\label{qce4}
\end{equation}

Notice that for quadratic expressions of the u$(3)$ generators, for the totally symmetric representation one has
\begin{eqnarray}
\bm{A}_{ij} \, \bm{A}_{k \ell} &=& \bm{b}^\dagger_i \, \bm{b}_j\,  \, \bm{b}^\dagger_k \, \bm{b}_\ell \,  \nonumber \\
&=& \bm{b}^\dagger_i \, \bm{b}^\dagger_k \, \bm{b}_j\,   \, \bm{b}_\ell + \delta_{jk} \, \bm{b}^\dagger_i \, \bm{b}_\ell \, . \nonumber
\end{eqnarray}
By means of this last expression one can define the two-body operator
\begin{equation*}
\hat{\cal O}(ik,j\ell) = \bm{A}_{ij} \, \bm{A}_{k \ell}  -\delta_{jk} \, \bm{A}_{i \ell} \, ,
\end{equation*}
and the expectation value of the two-body operator in the $N_a$-particle space can be written in the form
\begin{equation}
\langle \hat{\cal O}(ik,j\ell) \rangle^{(N_a)} = \frac{N_a!}{(N_a-2)! \, 2!} \, \langle \hat{\cal O}(ik,j\ell) \rangle^{(2)} \, .
\end{equation}
This expression can be used to determine the two-particle reduced density matrix by taking the expectation value of the quadratic expressions of the u$(3)$ generators with respect to the density matrix of $N_a$ particles. The reduced two-atom density matrix will be studied in a future contribution, here we shall study the one-atom reduced density matrix.

\bigskip

\subsection*{Entanglement}

\noindent
For a pure composite system of two subsystems, in our case matter and field, one has $S(\rho_{M\, F})=0$ and then, by means of the Araki-Lieb inequality,\footnote{$|S(\rho_M) - S(\rho_F)| \leq S(\rho_{M\, F}) \leq S(\rho_M) + S(\rho_F)$.} it is concluded that the mutual information is maximum and the von Neumann entropy of the matter is equal to the entropy of the field, $S(\rho_M)=S(\rho_F)$~\cite{nielsen11}.
Therefore the entanglement between the matter and field states of the Hamiltonian model may be calculated by means of the linear entropy $S_{\rm L} = 1-{\rm Tr}\, {\bm{\rho}^2_M} = 1-{\rm Tr}\, {\bm{\rho}^2_F}$ or the von Neumann entropy $S_{\rm VN} = - {\rm Tr}\, \bm{\rho}_M\, \ln \bm{\rho}_M = - {\rm Tr}\, \bm{\rho}_F\, \ln \bm{\rho}_F$, where $\bm{\rho}_\ell \, (\ell=Matter,Field)$ stands for the reduced density operator of subsystem $\ell$. Here we use the reduced density matrix of the matter to determine the entanglement between the matter and field sectors because its dimensions for $N_a$ particles is $d_M=(N_a +1)(N_a+2)/2$.

Note that the weight generators commute, whereas the lowering and raising generators change one atom from one energy level to another, so that their corresponding expectation values are zero for both, the full Hamiltonian and its RWA approximation, as they change the parity of the state or the value of the constants of motion $M$ and/or $K$. Therefore, for the exact quantum and symmetry-adapted-variational calculations, the one-atom reduced density matrix is diagonal, given only in terms of the occupation probabilities of the three-level system. The linear and von Neumann entropies for the variational symmetry-adapted and exact calculations in terms of the occupation probabilities, are given by
\begin{equation}
S_{\rm L} = 2 \sum^3_{j \neq k=1} {\cal P}_j \,  {\cal P}_k \,  , \qquad  S_{\rm VN} = -\sum^3_{j=1}  {\cal P}_j \ln {\cal P}_j\, .
\label{oneatom}
\end{equation}

In contrast to the previous result, for the coherent-state variational solution, being a tensorial product of matter and field coherent states, its linear and von Neumann entropies vanish, implying that there are no quantum correlations between the one-particle reduced density matrix and the rest of the system. This is true for any of the atomic configurations. From Table~\ref{twfe}, it is straightforward to construct the one-atom reduced density matrix and establish the purity condition,
\begin{equation*}
\vert x\vert^2+\vert y\vert^2  = \langle \bm{A}_{11}\rangle\,  \langle \bm{A}_{22}\rangle+\langle \bm{A}_{11}\rangle \langle \bm{A}_{33}\rangle+\langle \bm{A}_{22}\rangle\, \langle \bm{A}_{33}\rangle \, ,
\end{equation*}
where 
\begin{eqnarray*}
x&=&\langle\bm{A}_{12}\rangle\, , \quad y=\langle\bm{A}_{13}\rangle \quad  \hbox {for V }\, ,\\
x&=&\langle\bm{A}_{13}\rangle \, , \quad  y=\langle\bm{A}_{23}\rangle\quad  \hbox {for $\Lambda$} \, ,\\
x&=&\langle\bm{A}_{12}\rangle \, , \quad y=\langle\bm{A}_{23}\rangle \quad \hbox {for $\Xi$ } \, . %
\end{eqnarray*}

\begin{figure*}
\begin{center}
\includegraphics[width=0.45\linewidth]{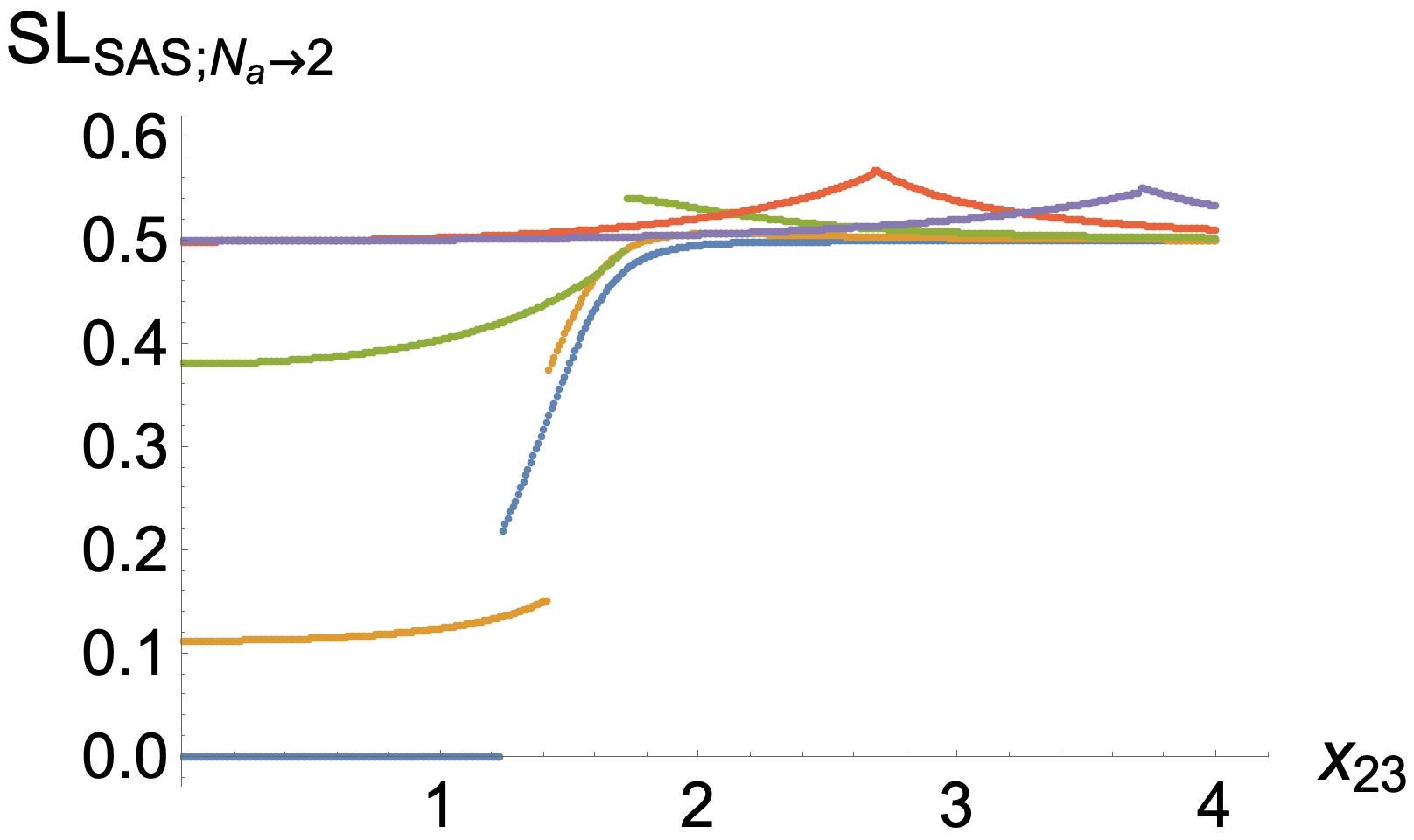}\qquad
\includegraphics[width=0.45\linewidth]{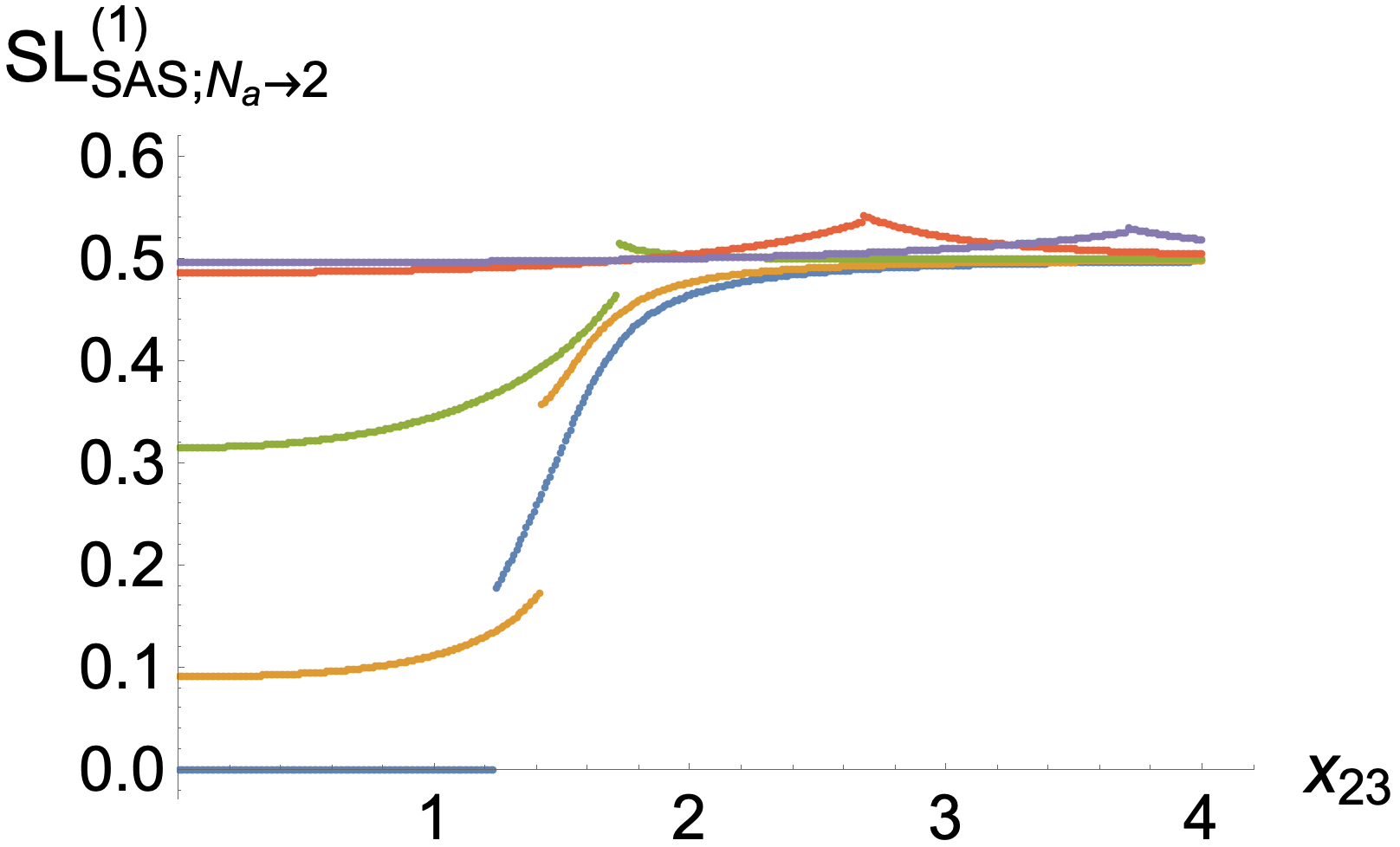}\\[3mm]
\includegraphics[width=0.45\linewidth]{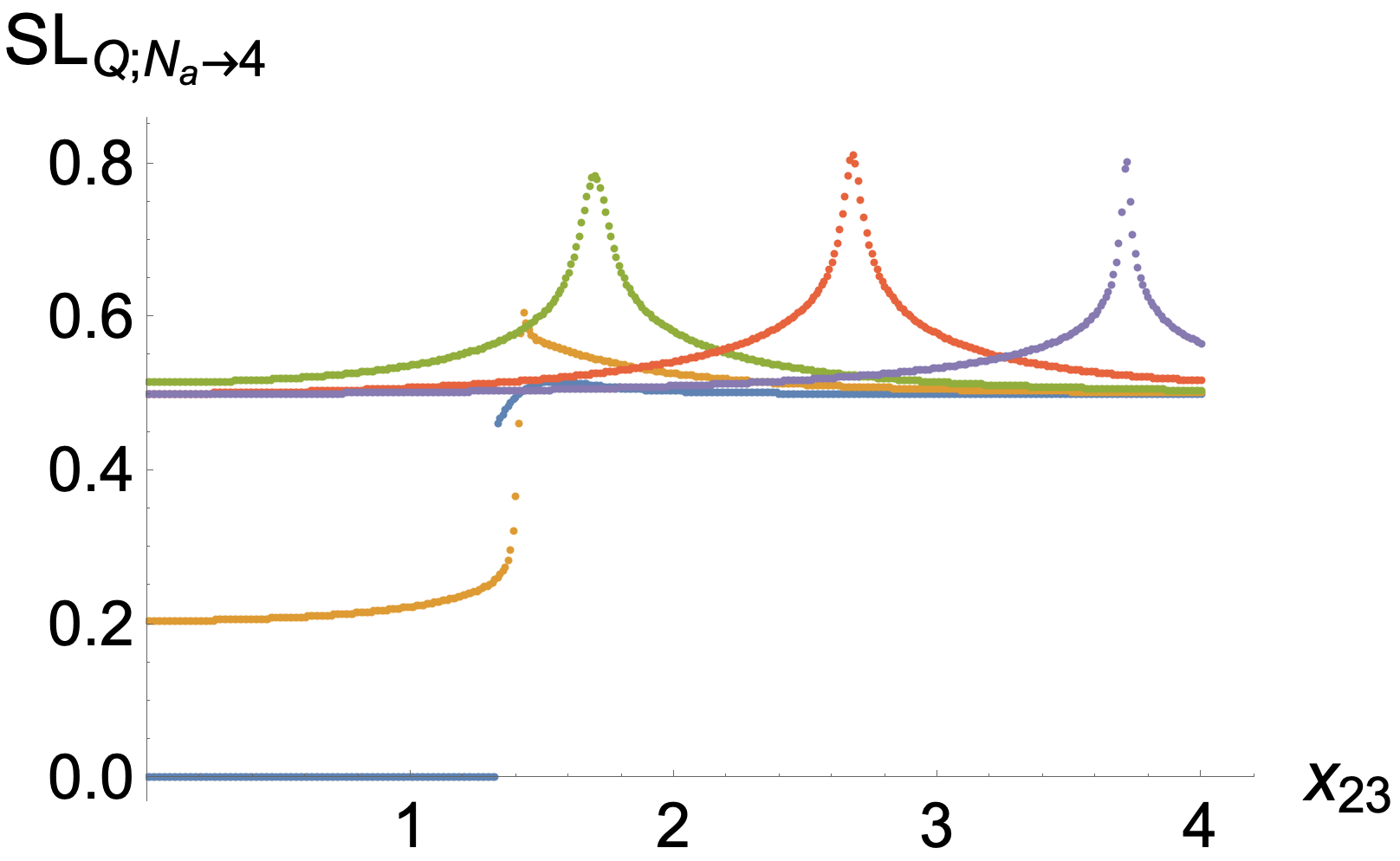}\qquad
\includegraphics[width=0.45\linewidth]{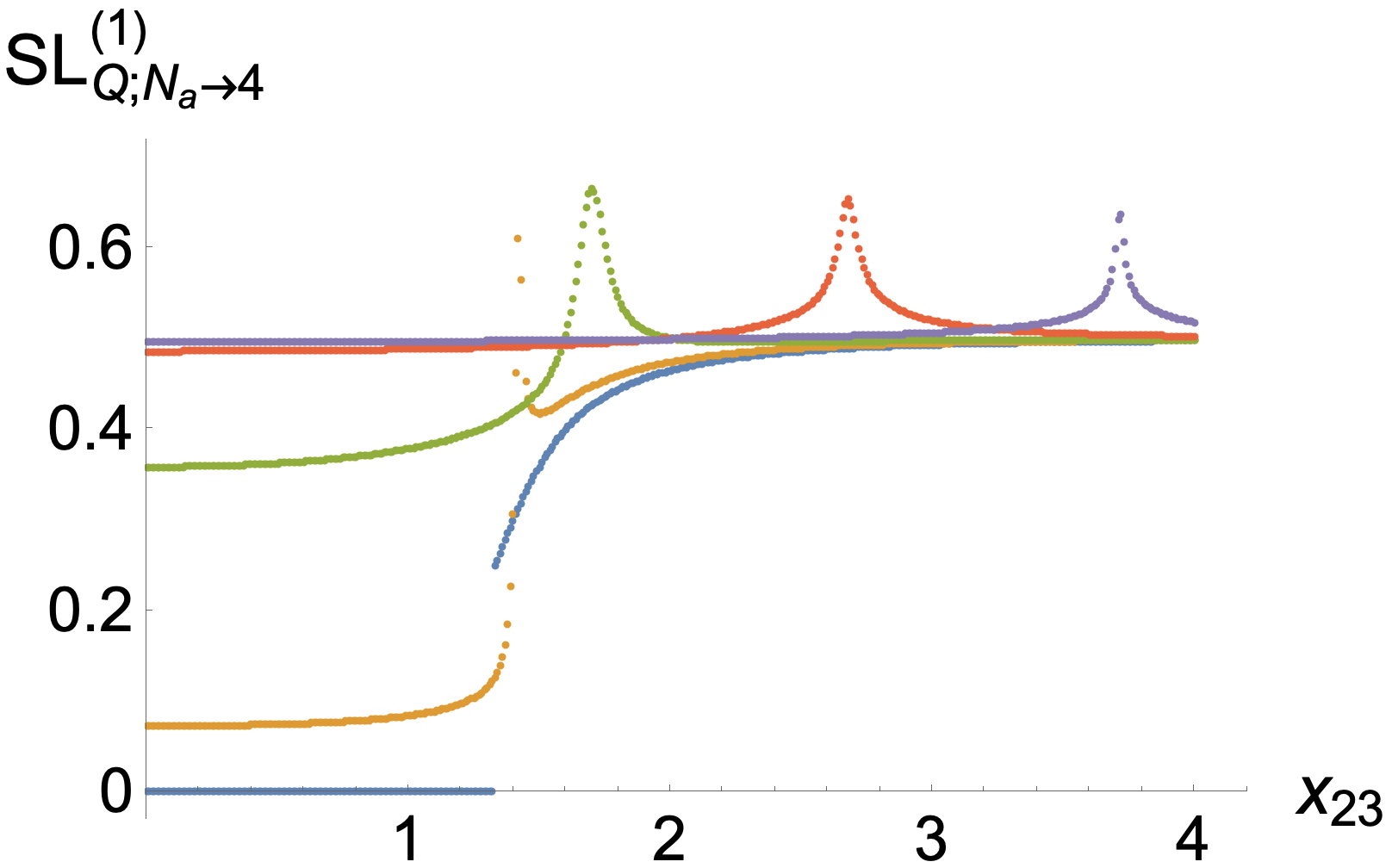}
\caption{Entanglement between field and matter (left), and the entanglement from one atom with the rest of the subsystem  (right), at top for SAS calculation with $N_a=2$  and at bottom for quantum exact calculation with $N_a=4$. The linear entropy is plotted against the dimensionless coupling strength $x_{23}$ for the following values of $x_{13}$, from bottom to top for each row: $x_{13}=0.1\ \textrm{(blue)},\ 1.0\ \textrm{(gold)},\ 1.5\ \textrm{(green)},\ 2.5\ \textrm{(red)},\ 3.5\ \textrm{(purple)}$. The singular points in the plots determine the transition loci. The entanglement between field and matter, and that between one atom and the rest of the system, produce the same critical curves.}
\label{f.entropy}
\end{center}
\end{figure*}

The calculations of the matter-field entanglement from the expression for $S_{\rm L}$ are shown in Figure~\ref{f.entropy}(left). The first row shows the results for the SAS ground state  with $N_a=2$, and the second for the exact quantum ground state with ${N_a=4}$ particles. They are compared with those on the corresponding right hand side obtained from the expression~(\ref{oneatom}). The plots are given as a function of $x_{23}$ for different values of $x_{13}$. All the results displayed maxima  of the entanglement measure at the transition points.
The comparison between the matter-field entanglement and that of one-atom with the rest of the system tells us about how much information they have in common, through the mutual quantum information function defined by~\cite{nielsen11}
\begin{equation*}
I(\rho_{n_1}:\rho_{n_2}) =S^{(n_1)}_L + S^{(n_2)}_L - S_L(\rho_{\rm M}) \, ,
\end{equation*}
where $\rho_{n_i}$ is the reduced density matrix of the matter subsystem with $n_i$ atoms. In particular for our case of two atoms, we have 
$
I(\rho_1:\rho_1) = 2 S^{(1)}_L -S_L( \rho_{\rm M}) \, .
$
Additionally, for the quantum mutual information between the matter and field sectors one has $ I(\rho_M:\rho_F) = 2 S(\rho_M)$ and it takes a maximum value as was mentioned above.

\subsection*{Simplex representation}

\noindent
The geometry of quantum states has become a relevant subject because it has potential applications in quantum information theory.  In particular, it tries to understand the properties of the space of states. Here we consider the probability distributions associated to the one-atom reduced density matrix of the considered Hamiltonian system, which can be interpreted as a qutrit. The set of all the eigenvalues form a two-dimensional simplex~\cite{bengtsson17}, 
\begin{equation*}
\bm{p}=  {\cal P}_1 \, \hat{i} +  {\cal P}_2 \, \hat{j} + {\cal P}_3  \, \hat{k} \, , 
\end{equation*}
with $\sum_k \, {\cal P}_k=1$ and ${\cal P}_k\geq0$. This set of points can be represented geometrically by an equilateral triangle of side length $\sqrt{2}$ in the space of occupation probabilities $\mathcal{P}_k$. The geometrical center of the equilateral triangle is given by the point $(1/3,1/3,1/3)$, The conditions
\begin{equation}
\left(\mathcal{P}_1-\frac{1}{3}\right)^2 + \left(\mathcal{P}_2-\frac{1}{3}\right)^2+ \left(\mathcal{P}_3-\frac{1}{3}\right)^2 = \frac{1}{6} \, , \qquad \mathcal{P}_1+ \mathcal{P}_2+\mathcal{P}_3=1 \, ,
\end{equation}
describe the largest inscribed circumference in the equilateral triangle, and the conservation of the total occupation probability, respectively. From these one calculates
\begin{eqnarray}
\mathcal{P}_1 &=& \frac{1}{2} \left( 1-\mathcal{P}_3 \pm \sqrt{\mathcal{P}_3 \, (2- 3 \, \mathcal{P}_3)}\right) \, , \quad 
\mathcal{P}_2 = \frac{1}{2} \left( 1-\mathcal{P}_3 \mp \sqrt{\mathcal{P}_3 \, (2- 3\, \mathcal{P}_3)}\right) \, ,
\end{eqnarray}
which allow us to determine the linear entropy of a qutrit on the inscribed circumference, as,
$S_L = \frac{1}{2}$ 
i.e., all the points on the circumference have the same linear entropy.

\begin{figure}[h!]
\begin{center}
\includegraphics[width=0.45\linewidth]{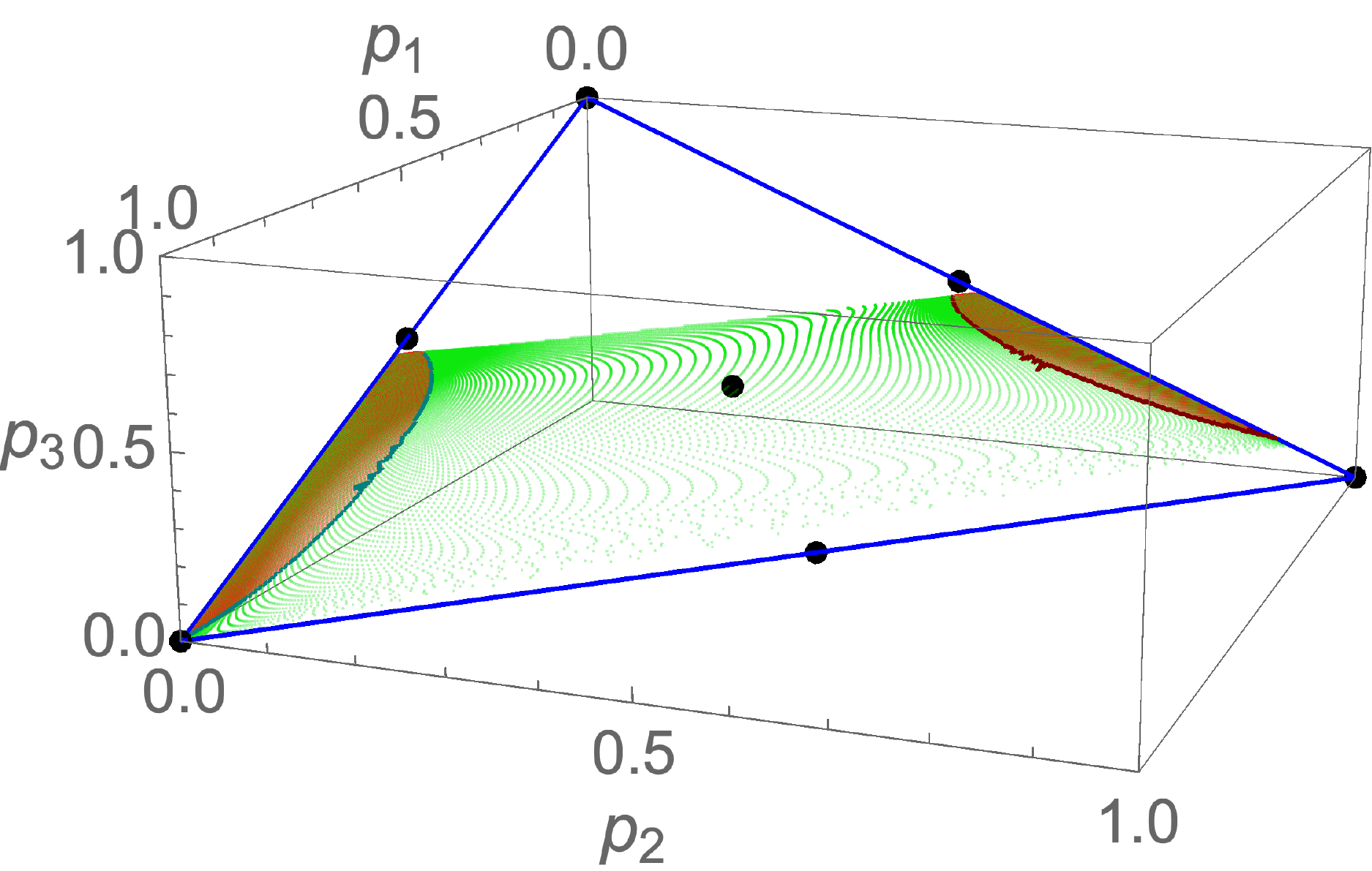}\qquad
\includegraphics[width=0.45\linewidth]{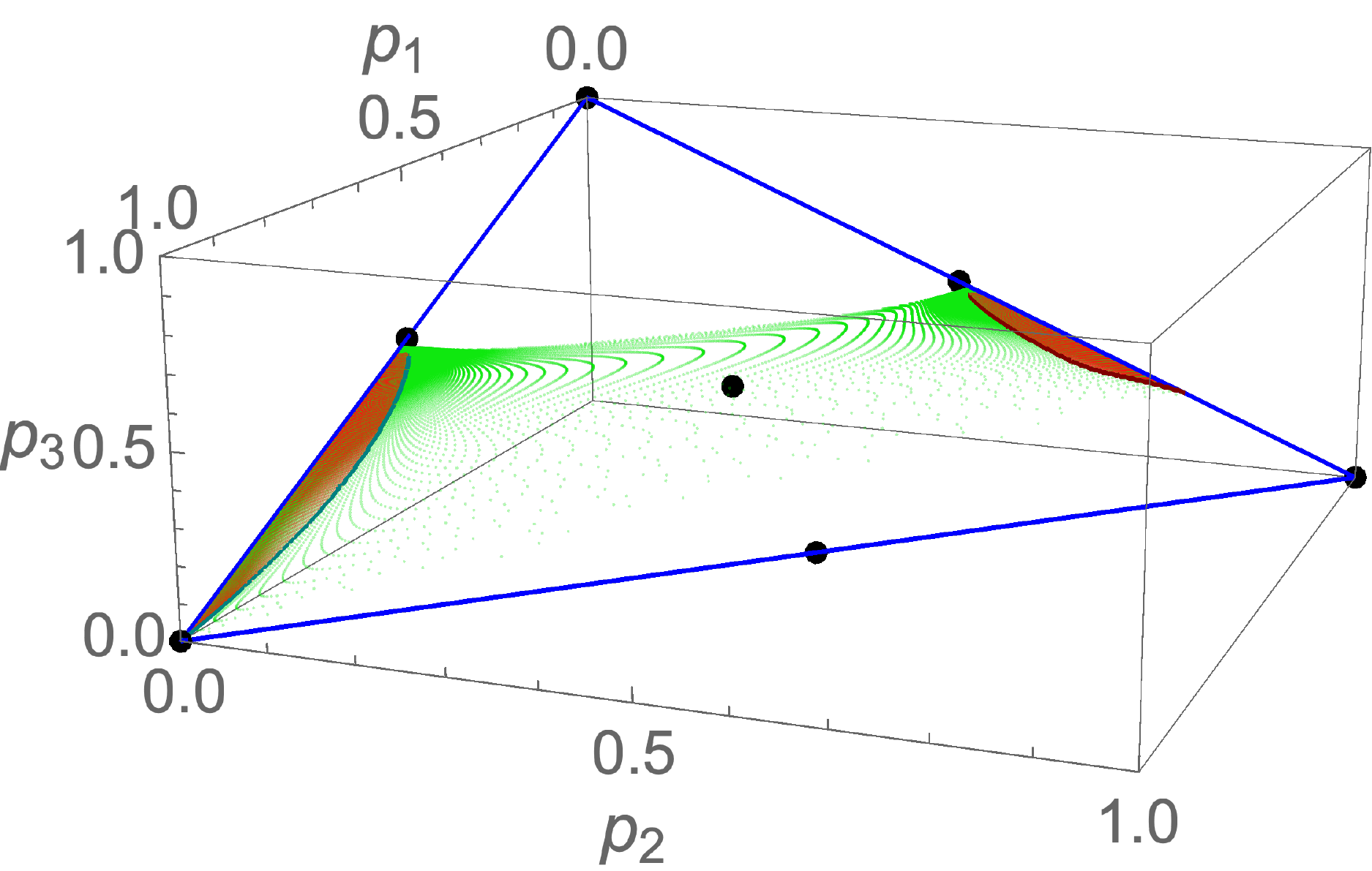}
\caption{Simplex representation of the quantum and SAS-variational solutions of the ground state for $N_a=2$ (left) and $N_a=4$ (right) atoms. The SAS results are indicated by (red) dots close to the edges of the triangle. The quantum results are indicated by (green) dots across the surface of the simplex. Notice that for an increasing number of particles, the simplex representation goes to the variational result, that is, all the points fall along the edges of the triangle}
\label{simplexLambda}
\end{center}
\end{figure}

The corresponding simplexes for $N_a=2$ (left) and $N_a=4$ (right) are displayed in Fig.~{\ref{simplexLambda} by the distribution of points on the surface bounded by the triangle (continuous blue line). In these figures, we have indicated with black dots the vertices where the states of the system are pure, and the middle points along the edges of the simplex where the subsystems $\mathcal{S}_{12}$, $\mathcal{S}_{13}$, and $\mathcal{S}_{23}$ have equal occupation probabilities; finally, the black dot at the point $(1/3,1/3,1/3)$ indicates the state with maximum mixture and entanglement. This triangle gives an indication of the purity of the one-particle density matrix; however, as the density matrix is diagonal, one obtains also information about its entanglement properties with the rest of the system. The (red) dots close to the edges of the triangle correspond to SAS variational results, obtained by a representative sampling of all the values of the coupling strengths $x_{13}$ and $x_{23}$; the border of these regions are determined by the separatrix in the collective region. The (green) dots across the surface of the simplex correspond to the quantum solution obtained from the diagonalisation of the Hamiltonian at the same representative points of the coupling strengths.

One notices, in both cases, that the quantum calculation gives states with large values of entanglement which the SAS results do not reproduce completely, in spite of the fact that they have the same symmetry properties, i.e., they all belong to the even-even sector of the Hilbert space, where the ground state lives. The density of points away from the edges is greater for the $N_a=2$ case than for $N_a=4$, a result which is expected because in the limit $N_a \to \infty$ all dots must tend to the edges of the triangle, since the energy surface of the SAS and variational coherent calculations take the same analytical form. 

\begin{figure*}
\begin{center}
\includegraphics[width=0.44\linewidth]{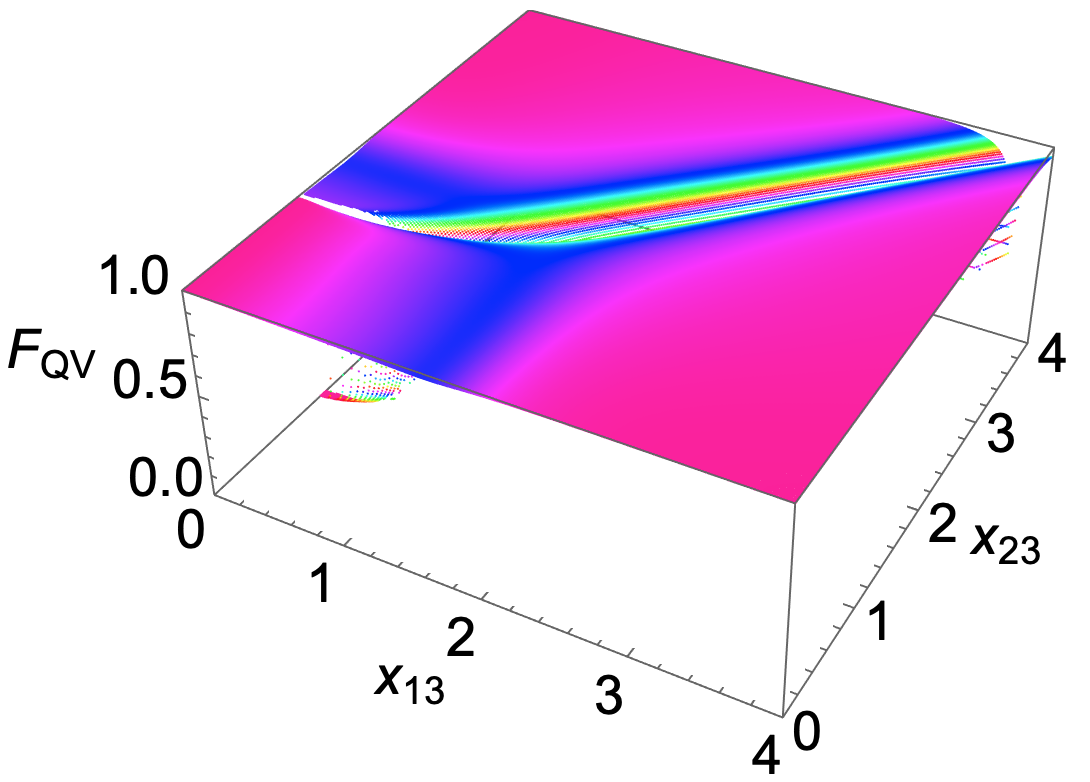} \quad
\includegraphics[width=0.44\linewidth]{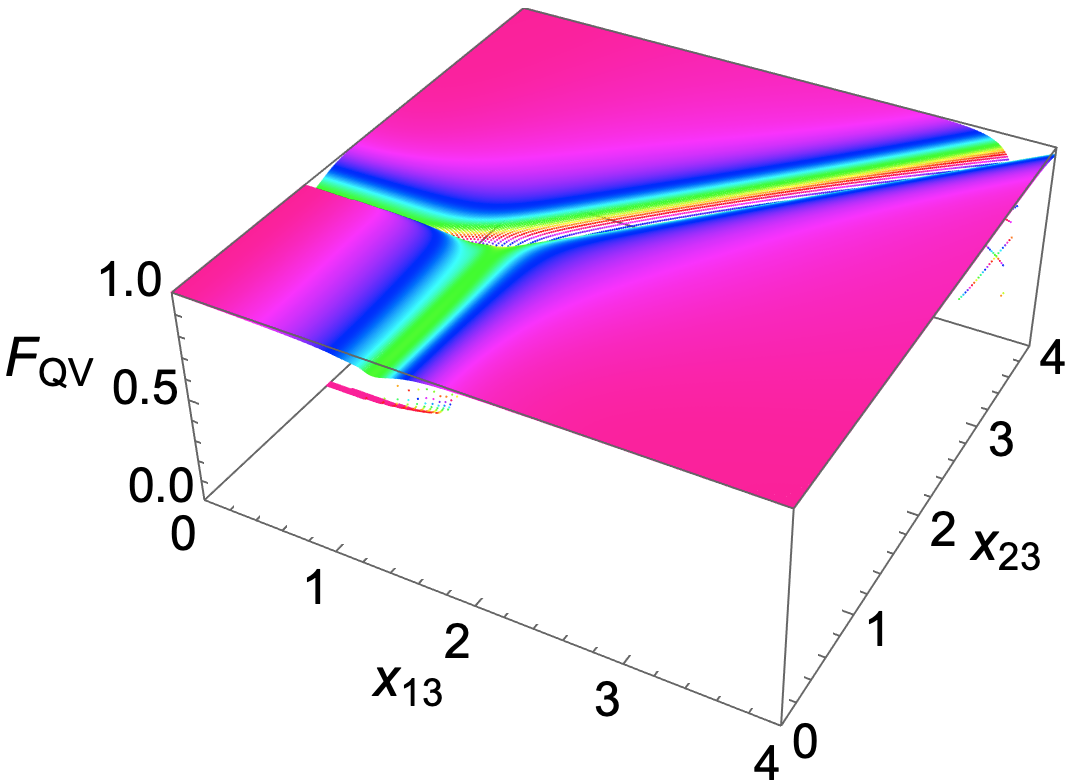}
\caption{Fidelity between the SAS state and exact quantum solutions for $N_a=2$ (left) and $N_a=4$ (right).  The transitions $N\rightleftharpoons S_{23}$ and $ S_{13}\rightleftharpoons S_{23}$ are of first order, and the fidelity goes to zero when $N_a\to \infty$. Observe that for the second order transition $N\rightleftharpoons S_{13}$ the fidelity is always different from zero.} 
\label{f.fidelidadQSAS}
\end{center}
\end{figure*}

To understand the difference obtained in the simplex representation of the qutrit for $N_a=2$ and $N_a=4$, we use again the fidelity concept from quantum information. Fig.~\ref{f.fidelidadQSAS} shows the fidelity between the reduced density matrices of the matter sector associated to the SAS and the quantum solution. We find that the two reduced density matrices are equivalent in most of the control parameter space, taking the value 1. However, significant differences appear in the vicinity of the separatrix where the fidelity takes values different from 1, which is narrower for $N_a=4$ than for $N_a=2$. In the first order transitions of the separatrix the fidelity between states drops almost to zero.

Thus, one may guess that the large quantity of points with greater entanglement and mixture appearing in~Fig.~\ref{simplexLambda} can be associated to the points in the neighbourhood of the separatrix. 

This conjecture can be proved by considering the simplex representation of a qutrit in Fig.~\ref{simplexLambda2}. The locus of maximum mixture at the same time indicates the state with maximum entanglement between one atom and the rest of the system. Additionally, we have proved using $S_L=\frac{1}{2}$ that all the points in the simplex representation that are at the same radius from the point $(1/3,1/3,1/3)$ have the same purity value or, equivalently, the same entanglement.  The radius of the circumference inscribed in the equilateral triangle is $r=\frac{1}{\sqrt 6}$, whereby one can conclude that the purity $P$ lies within the interval $P \in [1/3,\,1/2]$ or, equivalently, the linear entropy is in the interval $1/2\leq S_L\leq 2/3$ at Figs.~\ref{simplexLambda} and \ref{simplexLambda2}.

\begin{figure}[h!]
\begin{center}
\includegraphics[width=0.4\linewidth]{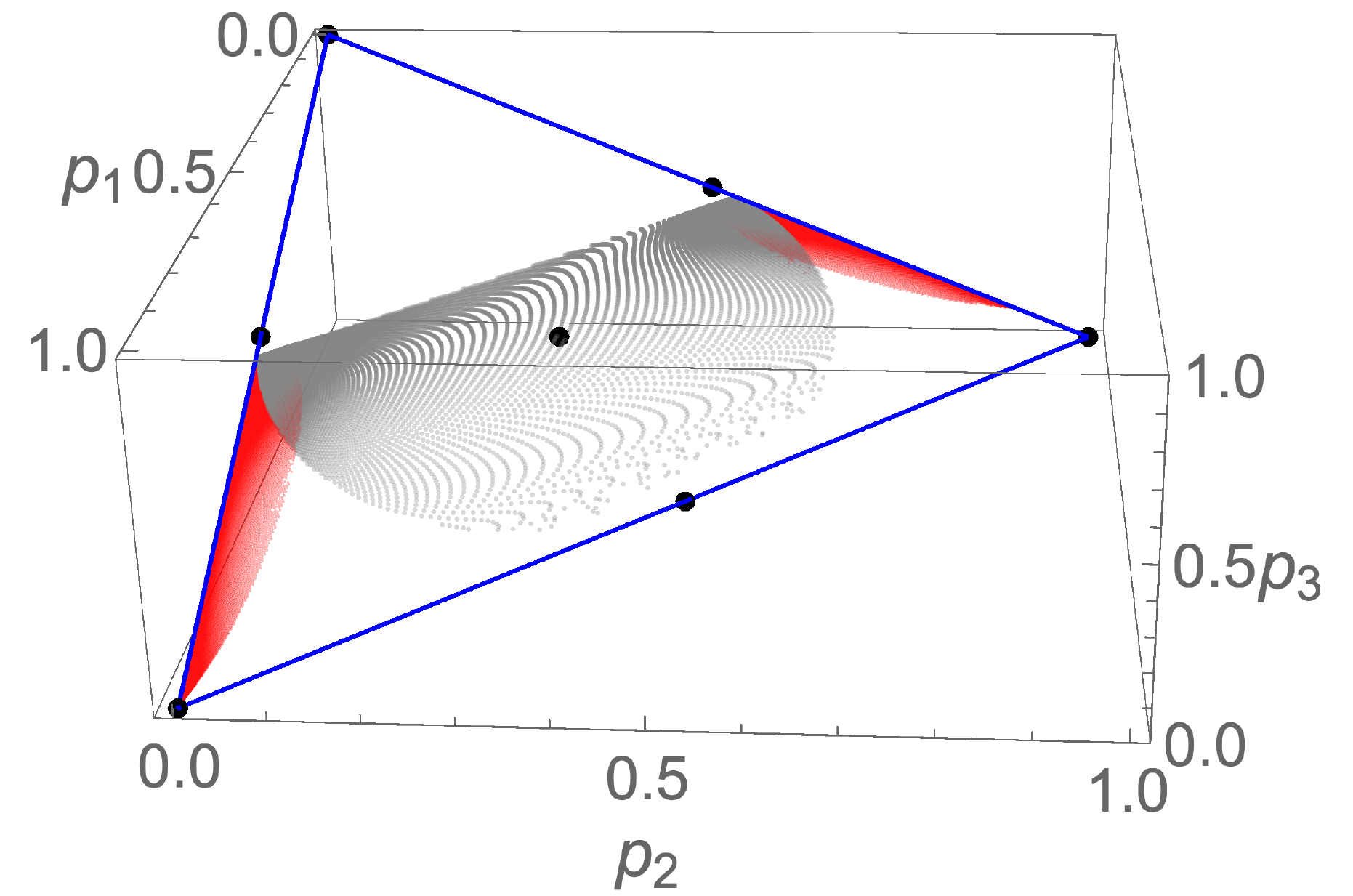}\quad
\includegraphics[width=0.35\linewidth]{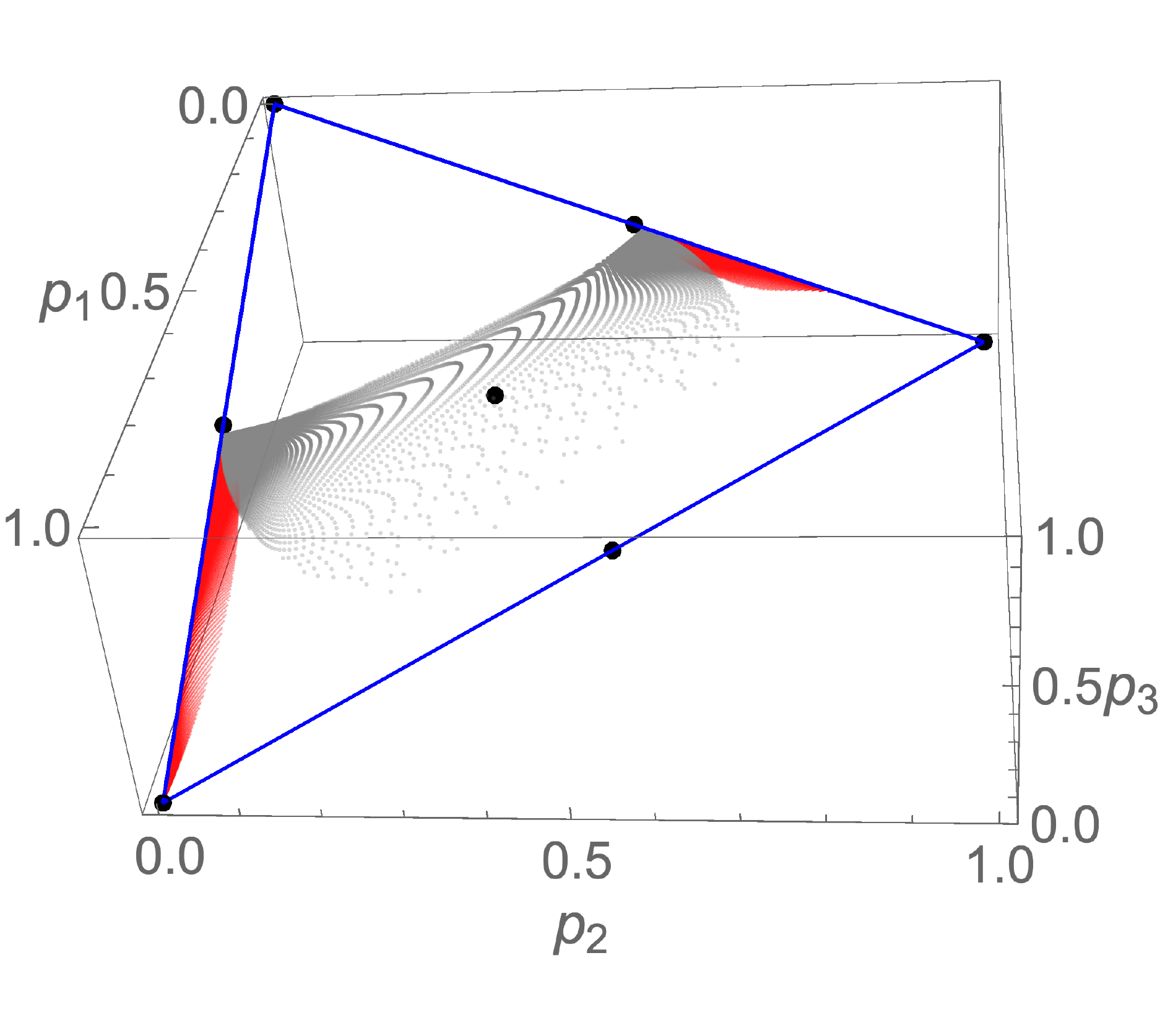}
\caption{Simplex representation of the occupation probabilities of the three-level system for $N_a=2$ (left) and $N_a=4$ (right). For the case $N_a=4$, the density of points with entanglement in the interval $1/2\leq S_L\leq 2/3$ (semicircular region) is smaller than that for $N_a=2$. The simplex representation gives information of the entanglement when the reduced density matrix of the matter sector is diagonal.}
\label{simplexLambda2}
\end{center}
\end{figure}

Therefore, the difference between the SAS and quantum states takes place in the vicinity of the separatrix, where, at these points, they exhibit the largest entanglement properties. This is also true for the entanglement between the matter and field.

\section{Summary and conclusions}
\label{s.conclusions}

A review of the quantum phase transitions was made for three-level atoms interacting with two modes of an electromagnetic field in a cavity, in the limit when $N_a$ goes to infinity.  Analytic expressions for the ground state energies, their eigenstates, and the corresponding separatrices were obtained.

For the variational SAS states, the energy surface of the ground state was obtained numerically, because the critical points cannot be determined in analytic form. Using the SAS ground state we construct the reduced density matrix of the matter sector, and calculate the maximum fidelity susceptibility surface. This surface is shown in Fig.~\ref{qpBures}, whose maximum values provide the phase diagrams for $N_a=2$ and $N_a=4$ cases, which closely resembles the quantum phase diagram of the $\Lambda$-atomic configuration in the limit $N_a \to \infty$. 

The entanglement between the matter and field sectors was calculated for the SAS and exact quantum calculation, for $N_a=2$ and $N_a=4$, for several values of $x_{13}$ and compared with the entanglement of one atom with the rest of the system. One sees (cf. Fig.~\ref{f.entropy}) similar results for the maximum values of the linear entropy identifying the transition points.

The fidelity between the SAS and exact quantum ground states is displayed in~Fig.~\ref{f.fidelidadQSAS}. One notices that the discrepancies appear in the vicinity of the separatrix.  which is narrower for $N_a=4$ than for $N_a=2$. We should mention that at the first order transitions which appears between the sectors ${\cal S}_{13}$ and ${\cal S}_{23}$ the fidelity goes to zero for $N_a\to \infty$.

 The quantum correlations between the matter and field were determined together with those associated to one atom and the rest of the system (composed by atoms and modes of the electromagnetic field) and are displayed in Fig.~\ref{simplexLambda}.  For the reduced one-atom density matrix of the ground state a simplex representation was given, which is very useful in order to see geometrically the mixture and entanglement of one atom with the system. Additionally we have established that the geometric differences between the exact numerical calculation and the SAS results are points in the parameter space with entanglement measures in the range $1/2 \leq S_L\leq 2/3$ (Fig.~\ref{simplexLambda2}).  
 
 \section*{Acknowledgments}
 
This work was partially supported by DGAPA-UNAM (under projects IN112520, and IN100323).

\section*{References}



\providecommand{\newblock}{}

\end{document}